\documentclass[aps,reprint,pre,superscriptaddress]{revtex4-2}
\usepackage{graphicx}
\usepackage{dcolumn}
\usepackage{amsmath,amssymb}
\usepackage{dcolumn}
\usepackage{bm}
\usepackage{xcolor}
\usepackage{color}
\usepackage[utf8]{inputenc}
\usepackage[T1]{fontenc}
\usepackage{mathptmx}
\usepackage{etoolbox}
\usepackage{textcomp} 
\usepackage{enumitem}
\usepackage{booktabs}
\makeatletter
\def\@email#1#2{%
 \endgroup
 \patchcmd{\titleblock@produce}
  {\frontmatter@RRAPformat}
  {\frontmatter@RRAPformat{\produce@RRAP{*#1\href{mailto:#2}{#2}}}\frontmatter@RRAPformat}
  {}{}
}%
\makeatother
\usepackage{hyperref}
\hypersetup{colorlinks,linkcolor={blue},citecolor={blue},urlcolor={blue}}

\date{\today}
\begin{document}

\title{Geometric Brownian information engine with finite cycle time: Optimisation of output work, power and efficiency} 

\author{Syed Yunus Ali}
\affiliation{Department of Chemistry, Indian Institute of Technology Tirupati, Yerpedu 517619, Andhra Pradesh, India}
\affiliation{Department of Physics, Indian Institute of Technology Bombay, Pawai, Maharastha, India}
\author{Rafna Rafeek}
\affiliation{Department of Chemistry, Indian Institute of Technology Tirupati, Yerpedu 517619, Andhra Pradesh, India}
\author{Debasish Mondal}
\email{debasish@iittp.ac.in}
\affiliation{Department of Chemistry,  Indian Institute of Technology Tirupati, Yerpedu 517619, Andhra Pradesh, India}

\date{\today}

\begin{abstract}
We consider a Geometric Brownian Information Engine to explore the effects of finite cycle time $(\tau)$ on the extractable work, power, and efficiency.  We incorporate an error-free feedback controller that converts the information obtained about the state of overdamped Brownian particles, confined within a 2-D monolobal geometry,  into extractable work.  The performance of the information engine depends on the cycle period $(\tau)$, measurement distance $(x_m)$, and feedback location $(x_f)$ of the controller.  Upon increasing the feedback cycle time, the engine transitions from a high non-equilibrium steady state to a completely relaxed state. We set the measurement distance at an optimum position related to a fully relaxed state ($x_m^* \sim 0.6 \sigma$).
When the cycle time is finite and short ($\tau <\tau_r$), the best information processing occurs with a shorter distance of the feedback site. While increasing the cycle time towards a fully relaxed state ($\tau  \gg \tau_r$), the maximum extractable work that can be achieved with a feedback location is set to be twice that of $x_m^*$, as expected.  When the cycle time ($\tau$) is longer than the relaxation time ($\tau_r$), the maximum power is achieved when the scaled feedback location is exactly double the optimum measurement distance ($x_f^{*}=2x_m^*$). In contrast, when $\tau < \tau_r$, the maximum power is achieved when the feedback site is set at a lower value. As the $\tau$ increases, the maximum average power decreases. In the limit of a long $\tau$, the highest efficiency as well extractable work is attained when $x_f$ is located at $2x_m$, regardless of the level of entropic control. As the dominance of entropic control increases, the extractable work and efficiency in the fully relaxed state decrease due to higher information loss during relaxation.


\end{abstract}

\maketitle

\section{\label{sec:level1}Introduction}

The interplay between the chemical or thermal noise and the acquired information is observed in different cellular and artificial nano-machines, where information is used to control the transport of physio-chemical processes at the single molecular level \cite{parrondo2015nat,Linke2021pnas,Horowitz2013prl}.
A Brownian information engine (BIE) is a vital prototype for understanding the physical principles of such processes. A BIE is a device that can extract mechanical work from a single heat reservoir by utilizing information related to the positional surprise of Brownian particles \cite{Ashida2014pre,Berut2012nat,Paneru2018prl} under consideration. In 1871, Maxwell first introduced the idea of sorting particles based on their average velocity in the presence of a single heat bath \cite{maxwell1871theory,Rex2003maxwell}. The separation process reduces the system's (isolated) entropy spontaneously, apparently violating the second law of thermodynamics. The puzzle can be resolved considering a direct link between the demon's information and thermodynamic entropy \cite{Rex2003maxwell,Szilard1929zphys,Brillouin1951jcp,Landaueribm1961,Bennett1982intjtphys}. Later, Sagawa and Ueda first expressed a quantitative relationship between information change and the thermodynamic work done \cite{Sagwa2008prl,Sagawa2009prl,Sagawa2010prl} during a suitable state change. The relation explains the balance between the decrease of thermodynamic entropy and the quantity of information received during a measurement process. The extractable work cannot exceed the difference between the free-energy change and the available information obtained during a measurement \cite{Sagwa2008prl,Sagawa2009prl,Sagawa2010prl}. Because of the significant progress in stochastic thermodynamics in the last two decades, \cite{jarzynski1997prl,sekimoto2010stochastic,seifert2012rpp,Esposito2012epl}, and connected fluctuation relations \cite{Harris2007jsm,Jarzynski2011arcmp,Horowitz2010pre}, various theoretical prototypes of an information engine whose initial state is in thermal equilibrium have been investigated for both classical \cite{Abreu2011epl,Berut2012nat,Ashida2014pre,Jun2014prl,Pal2014pre,Ashida2014pre,Paneru2018prl,paneru2022jpcl,Rafna2024jcp,Rafna2025arxiv1,saha2021pnas, Rafna2025arxiv2,Rafna2025jpcb}, and quantum systems \cite{Sagwa2008prl,Kim2011prl,Bruschi2015pre,Goold2016jphysA,Masuyama2018natcom,Naghilo2018prl}. Meanwhile, several technological advances have resulted in novel experimental approaches that have enabled the implementation of many information engines in electronic and Brownian systems \cite{Koski2014pnas,Koski2014prl,Koski2015prl,Kutvonen2016Scirep,Lopez2008prl,Toyabe2010natphys,Paneru2018prl,Berut2012nat}.\\

Recent theoretical studies explore the performance of an information engine with a finite cycle time and an arbitrary initial state \cite{Esposito2011epl,Still2012prl,Abreu2012prl,Bauer2012jphysa,Park2016pre}. Interestingly, the condition for the maximum power is different than the same for the best efficiency of a standard reversible engine \cite{Park2016pre}. Therefore, these studies are crucial in exploring the time evolution of several relevant observables, such as extracted work, power, and process efficiency. Paneru et al. \cite{Paneru2018pre} have recently performed an experimental investigation of a BIE. In this setup, the particle, confined in an optical energy trap, evolves from a non-equilibrium steady state to a fully relaxed one. The procedure involves measurement and subsequent feedback control repeated with a finite cycle period. The study examines the time evolution of achievable work, engine efficiency, and the essentials for the best power. Another recent issue \cite{Lee2018scirep} investigates the thermodynamics of an information-driven Brownian motor with a cycle period $\tau$ related to the characteristic relaxation time of the particle. One of the important concerns related to the time evolution of efficiency and power of a BIE lies in the continuous change in the particle's surprise (information) during the relaxation process. The change in surprise leads to the change in available accrued information, affecting the engine's power and efficiency as a function of the cycle time. \\

So far, all studies related to the temporal evolution of power and efficiency of a BIE have considered an external energy potential (like optical trapping) as confinement potential \cite{Lee2018scirep,Paneru2018pre,Paneru2020Natcommun,paneru2022jpcl}. However, an information engine with an entropic (geometric) potential \cite{Zwanzig1992jpc,Reguera2001pre,Reguera2006prl,Burada2009cpc} has yet to receive substantial attention. Once confined in an irregular channel, a Brownian particle feels an effective phase space-dependent (entropic) potential along the direction of particle transport \cite{Zwanzig1992jpc,Reguera2001pre,Reguera2006prl,Burada2007pre,Burada2008prl,Burada2009cpc,Mondal2010pre,Mondal2010jcp1,pagliara2014prl,Mondal2011jcp,Marchesoni2009pre,Mondal2010jcp2,Burada2008biosyst,Das2012jcp1,Mondal2011jcp,Burada2009epl,Das2012pre,Quan2008jcp,Das2012jcp2,Nayak2020jpcc,
yang2017pnas,Zhu2022prl,Marquet2002prl,pagliara2014epjs}. Brownian diffusion inside a small cavity or channel is crucial for inter-cellular transport processes. Examples cover nutrient flow (in or out) into the bloodstream, cross-membrane transport of ions or macromolecules, and signal transduction in the synapses, to mention a few \cite{Borromeo2006pre,Zhou2010jpcl,Licata2016bioj,pagliara2014epjs,Arango2021biophys,Holek2019cat,carusela2021fcdb,Radi1998crt,Ricciardi_2013_diffusion,Kleinfeld1998biochem, Berez2022jcp,Mondal2016jcp1,Mondal2016jcp2,Muthukumar2003jcp,Mangeat2020jcp}.
Apart from chemical, thermal, and energetic imbalances, all these micro-machinery encounter space-dependent entropic driving (or inhibition) while performing the related cellular event. The importance of entropic constraints in diffusion processes thus triggered the attention of researchers in the recent past to understand the underlying principle of these physio-chemical processes and to design small-scale machines \cite{Wang2021pccp,santamaria2020pccp}. Therefore, detailed studies on the design principles of information engines in the presence of such entropic potential and monitoring the time-dependent performance are of potential interest. \\
\begin{figure}[!h]
    \includegraphics[width=0.44\textwidth]{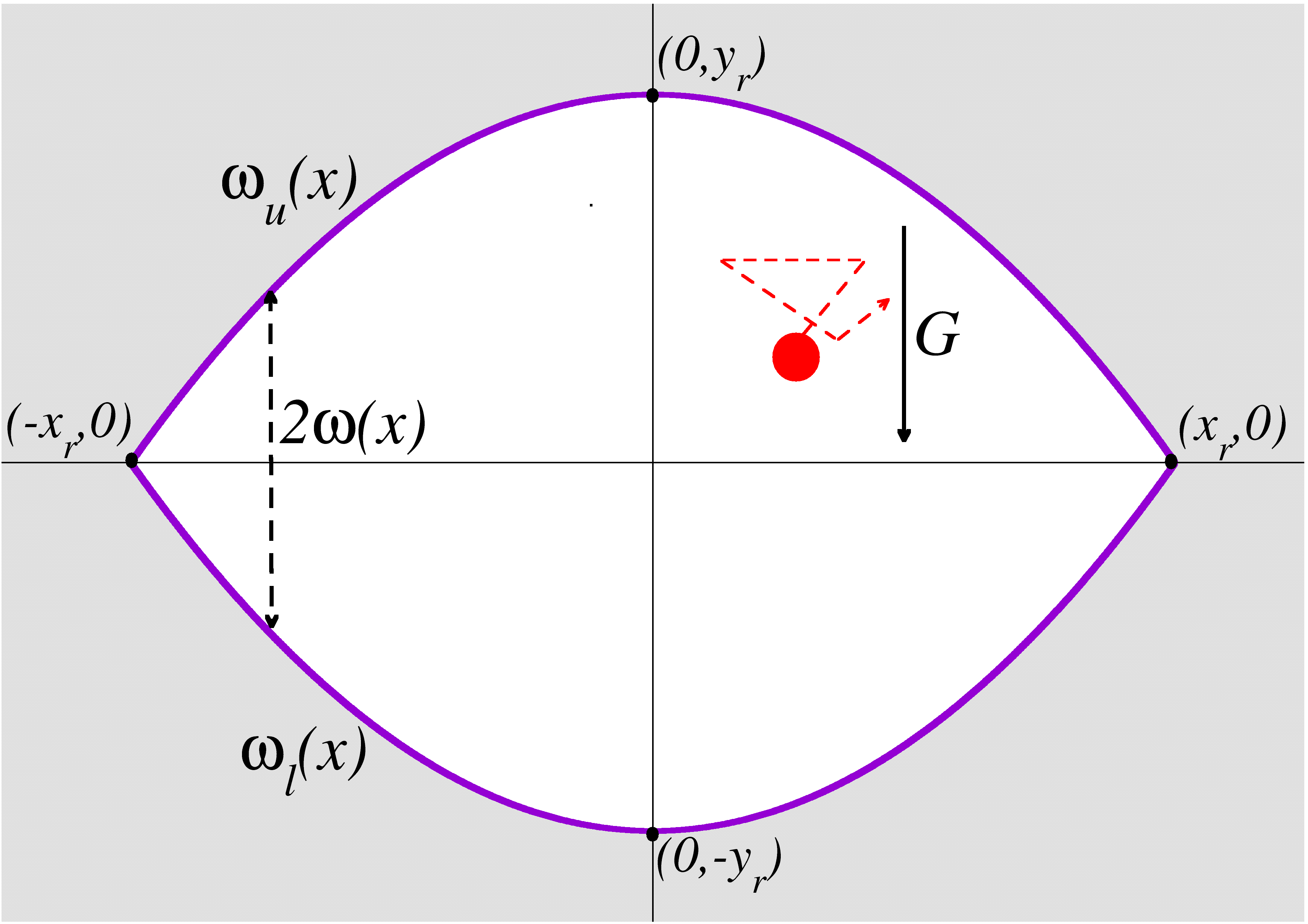}
    \caption{Schematic illustration of the confinement of Brownian particle in a two-dimensional monolobal confinement. $\omega_u$ and $\omega_l$ denote the upper boundary and the lower boundary of the confinement, respectively. $x_r$ and $y_r$ are characteristic length scales along the $x$ and $y$ directions, respectively. $\omega(x)$ is the local half-width at $x$.}
    \label{f1}
\end{figure}

Recently, we developed geometric Brownian information engine (GBIE) \cite{Ali2022jcp,Rafna2023pre} using overdamped Brownian particles contained within a two-dimensional monolobal confinement (Fig.~\ref{f1}). We employ an error-free feedback control protocol that comprises three stages: measurement, feedback, and relaxation. Outcomes of the information engine depend on the geometric constraints, the reference measurement length $x_m$, and feedback location $x_f$. We estimated the extractable work, complete information, and unavailable information associated with error-free feedback control using the equilibrium marginal probability distribution. We have determined the amount of available information that can be utilized in a long time limit ($\tau \rightarrow \infty$) and the optimum functioning requisites for best work extraction \cite{Rafna2023pre}. In the presence of symmetric feedback, we also pinpointed the precise value of the upper bound of extractable work as $( 5/3- 2\ln2)k_{B}T$ under a pure entropic dominance \cite{Ali2022jcp}.\\

The present study deals with the optimal tuning of the available information as an output of the GBIE under a non-equilibrium steady state condition with a finite cycle period time (finite $\tau$). We, thus, intend to calculate the average extractable work in the presence of a limited (finite) cycle time and the related power, i.e., averaged useful work per unit of time. 
We consider an error-free (almost) feedback scheme, similar to \cite{Paneru2018pre,Paneru2018prl,Paneru2020Natcommun,Ali2022jcp,Rafna2023pre,Rafna2024jcp}, that operates in a non-equilibrium steady state to a relaxed condition. We repeat the measurement and subsequent feedback control with a finite cycle period. We intend to examine how the requirement to get maximum extractable work evolves in time and its consequences on the power and efficiency of the machine. Finally, we address the role of entropic dominance in the time evolution of extractable work, power, and efficiency of the engine.
\section{MODEL AND METHOD}
\subsection{Dynamics of  Confined Brownian particles}
We consider an overdamped Brownian particle trapped inside a monolobal two-dimensional confinement \cite{Ali2022jcp,Rafna2023pre} as shown in Fig.~\ref{f1}. The length scale along the $x$ direction is much longer than the same along the perpendicular $y$ direction. A constant force $G$ is acting in the transverse direction. If $\vec{r}$ denotes the position of the particle in two dimensions, the Langevin equation
for the particle can be written as: 
\begin{equation}\label{eq1}
\begin{aligned}
   \frac{d \vec{r}}{dt} &= -G\hat{e_y} +\vec{\zeta}(t), \\
\end{aligned}
\end{equation}
where, $\vec{r}=x \hat{e_x} + y \hat{e_y}$. $\hat{e_x}$ and $\hat{e_y}$ are the unit  vector along the $x$ and $y$ directions, respectively. We have considered the frictional coefficient of the particle to be unity.
A Gaussian white noise $\vec{\zeta}(t)$ mimics the thermal fluctuations with the following properties:
\begin{equation}\label{eq2} 
\begin{aligned}
\left \langle \zeta_j (t)\right \rangle &= 0, \; &for \; j =x,y \\
\left \langle \zeta_i (t) \zeta_j ({t}')\right \rangle &= 2\beta^{-1}\delta_{ij}\delta (t-{t}'), \; & for\;i,j=x,y.
\end{aligned}
\end{equation}
Where, $\beta^{-1}=k_BT$ and $k_B$ is the Boltzmann Constant.  $T$ denotes the temperature in absolute scale. $\omega_u(x)$ and  $\omega_l(x)$ are the equations for the upper and lower walls,
respectively: 
\begin{equation}\label{eq3}
    \omega_u(x) =- \omega_l(x) = -ax^2+c.
\end{equation}
$a$ and $c$ are constant geometric parameters always positive. Consequently, the maximum length scale along the $y$ and $x$-directions are $2 y_r$ ($=2c$) and $2x_r$ ($=2\sqrt{c/a}$), respectively. The local width of the confinement at $x=x^{\prime}$ reads as $2\omega(x^{\prime}) = \omega_u(x^{\prime})-\omega_l(x^{\prime})$.\\

One can make the Eq.~\ref{eq1} dimensionless by scaling the lengths by $x_{r}$, temperature by a reference temperate $T_{R}$ and force by a reference force ($F_{r}$). If $p (x,y,t)$ describes the probability density of the particle at position $(x,y)$ at time $t$, Eq.~\ref{eq1} can alternatively be described by the following 2-D Fokker-Planck equation \cite{Burada2008prl,Reguera2001pre,Reguera2006prl,Burada2007pre,Burada2009epl,Mondal2010pre,Burada2008prl,Burada2009cpc, Quan2008jcp,Mondal2010jcp1,Burada2008biosyst,Das2012pre,Risken,GardiHSM,Burada2008prl,Das2012jcp1,Mondal2010jcp2,Mondal2011pre,Burada2009cpc}: 
\begin{align}\label{eq5}
\frac{\partial}{ \partial t} p(x, y, t) = & \beta^{-1}\frac{\partial}{\partial x} \left \{  e^{-\beta U(x,y)} \frac{\partial}{\partial x} e^{\beta U(x,y)}p(x,y,t)\right \} \\ \nonumber + & \beta^{-1}\frac{\partial}{\partial y}\left \{ e^{-\beta U(x,y)} \frac{\partial}{\partial y} e^{\beta U(x,y)}p(x,y,t)\right \},
\end{align}
with a potential function $U(x,y) = Gy$.  As the length scale along the longitudinal direction, $x_{r}$ is much larger than that of the transverse one ($y_{r}$), one can assume a fast local equilibrium along the $y$-direction \cite{Zwanzig1992jpc,Burada2008prl,Burada2009cpc,Reguera2001pre,Reguera2006prl,Burada2007pre,Mondal2010pre,Quan2008jcp,Mondal2010jcp1,Burada2008biosyst}. Under such assumption and if the spatial variation of a wall along the direction is not too high ($\left | {\omega}'(x) \right |\ll 1$), Eq.~\ref{eq5} can be reduced to a 1-D  Fokker-Planck description as  
\cite{Zwanzig1992jpc,Burada2008prl,Reguera2001pre,Reguera2006prl,Burada2007pre,Burada2009cpc,Mondal2010pre,Quan2008jcp,Mondal2010jcp1,Burada2008biosyst,Das2012jcp1,Burada2009epl,Das2012pre}: 
\begin{equation}\label{eq14}
    \frac{\partial}{ \partial t} P(x,t) = \frac{\partial}{\partial x}  \left \{ \beta^{-1}  \frac{\partial}{\partial x} P(x,t) +  A'(x)P(x,t) \right\}.
\end{equation}

Where $P(x,t)$ is a marginal probability density and can be obtained by integrating the occupation over the entire transverse coordinate at a given $x$:
\begin{equation}\label{eq6}
P(x,t) = \int_{y } p(x,y,t ) dy, 
\end{equation}
 and  $A(x)$  is the effective potential experienced by the particle along the direction of transport: 
  \begin{equation}\label{eq13}
    \begin{aligned}
    A(x) = -\beta^{-1} \ln \bigg[\frac{2}{G \beta} \sinh \bigg( \beta  G\omega(x) \bigg) \bigg].
    \end{aligned}
\end{equation}
 
The effective potential $A(x)$ does not exist in the actual 2-D Langevin dynamics but arises in reduced dimensions due to the entropic constraints associated with confinement. Eq.~(\ref{eq13}) yields $A(x) = - G\omega(x)$ for  $\beta G \omega(x) \gg 1$. In this limit, the transverse force dominates over thermal fluctuations, and particles can move only in close vicinity to the lower wall of the confinement. Hence, the reduced description becomes equivalent to a Brownian particle in a purely energetic trap \cite{Burada2008prl,Mondal2010pre}. In the other extreme $\beta G \omega(x) \ll 1$, the effective potential is independent of G: $A(x) = -\beta^{-1} \ln[2 \omega(x)]$. In this limit, a particle can easily avail the entire phase space of the confinement. We refer to this limit as entropy controlled situation \cite{Burada2008prl,Mondal2010pre}. The detailed derivation of Eq.~\ref{eq14} from Eq.~\ref{eq5} can be obtained in \cite{Zwanzig1992jpc,Reguera2001pre,Reguera2006prl,Burada2007pre,Burada2008prl,Burada2009cpc,Mondal2010pre,Mondal2010jcp1}.\\
 
Before we proceed, we mention two pertinent points. First, an effective logarithmic potential appears in other various entropy-driven biophysical processes, like polymer translocation \cite{Muthukumar2003jcp,Mondal2016jcp1,Mondal2016jcp2} DNA unzipping events \cite{Poland1966jcp1,Poland1966jcp2,Bar2007prl,Kaiser2014jphysa,Fogedby2007prl} and in the case of optically trapped cold atoms, as well \cite{Kessler2012prl,Kessler2010prl,Lutz2013nphys}. Second, recent experimental developments suggest that one can fabricate mesoscopic channels with irregular cross-sections along the transport direction \cite{Zhu2022prl,yang2017pnas,Marquet2002prl,pagliara2014epjs,pagliara2014prl}. For example, one may use a two-photon writing system followed by the imaging procedure to design such cavity \cite{yang2017pnas,Zhu2022prl}. Similarly, the photo-lithographic technique is also helpful in microfabricating a narrow channel with an irregular shape \cite{Marquet2002prl}. A careful engineering prototype of microfluidics and holographic optical tweezers can also mimic the constraint diffusion inside a small cavity \cite{pagliara2014epjs,pagliara2014prl}. 
\subsection{Feedback Protocol and Numerical methods}
\begin{figure}
    \centering
    \includegraphics[width=0.45\textwidth]{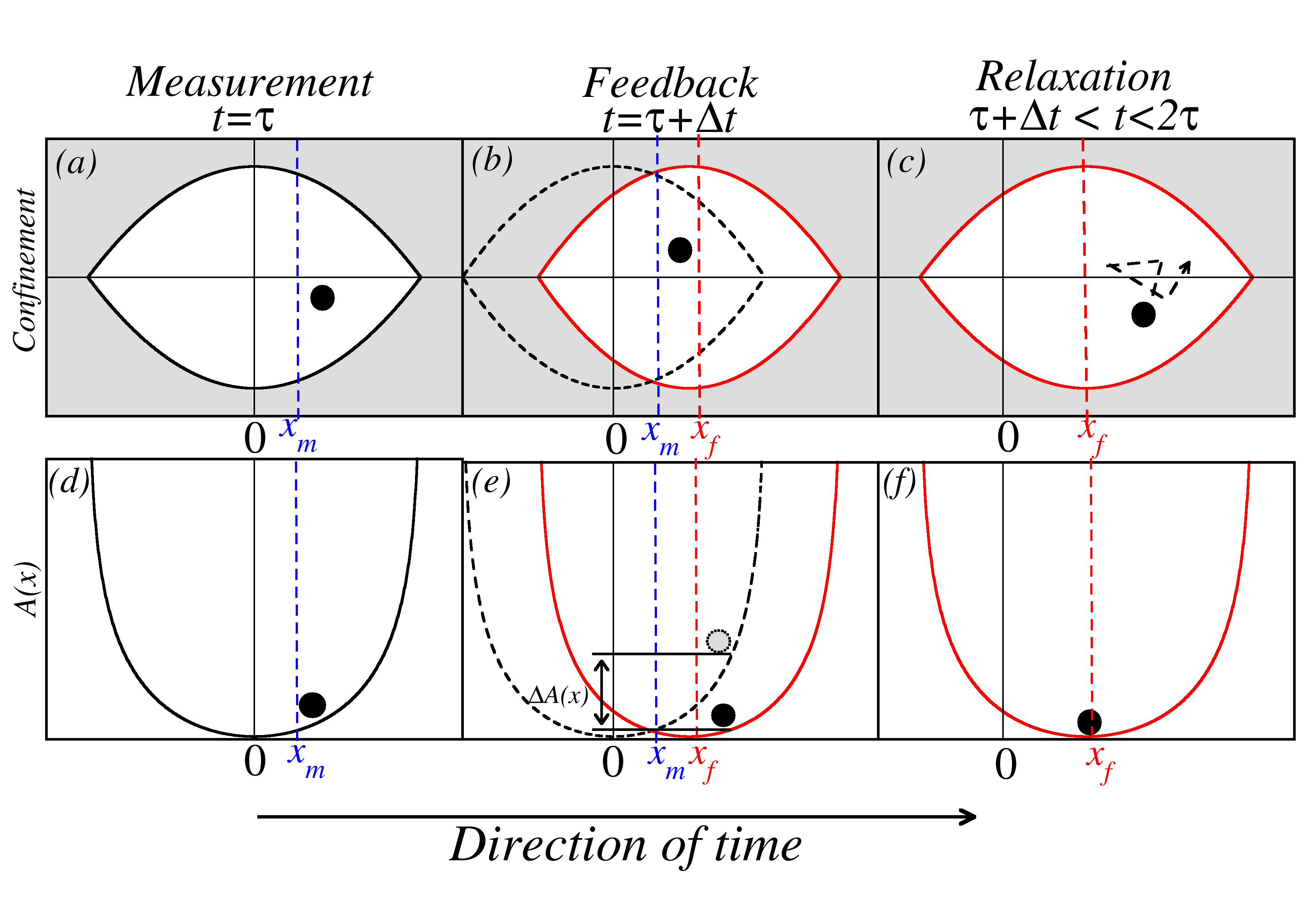}
    \caption{Model description of the feedback cycle. A particle is immersed in a monolobal confinement. Initially, the confinement centre has been set at the origin ($\lambda(t)=0$). First, we set a measurement site $x_m$ to identify the particle's location. No action is taken if the particle resides left to the $x_m$. However, if the particle stays right to $x_m$, we instantaneously shift the confinement center to $x_f$ $(\lambda(t)=0$ to $\lambda(t)=x_f)$. After the feedback step, the particle relaxes for $\tau$ time with a fixed trap centre $(x_f)$. We measure the physical observable at $t=\tau$ and repeat the cycle once again, starting with a fully relaxed state.}
    \label{f2}
\end{figure}
We consider that $\lambda(t)$ is the location of the confinement centre. Initially, $\lambda(0)$ is set as zero. One feedback cycle consists of three processes:  measurement, feedback, and relaxation for the time $\tau$.
Feedback scheme is illustrated in Fig.~\ref{f2}. First, we set a reference distance $x_m$, then measure the particle position of a fully relaxed state. We shift the confinement centre to $\lambda(t)=x_f$ if and only if $x\geq x_m$. Whereas for $x < x_m$, there has been no action done. We start calculating the relaxation time at the time of the feedback and allow the particle to relax up to $\tau$ time with a fixed trap center. We measure all required physical observables and repeat the cycle. 
In an effective reduced description, the particle experiences an effective potential $A(x-\lambda(t))$ \cite{Ali2022jcp,Rafna2023pre}.
Thus, the change in effective potential energy ($\Delta A(x)$), equivalent to the free energy change during the feedback, is translated into heat and work. However, the confinement and related potential are shifted instantaneously. Thus, the particle does not have time to travel and dissipate energy. As a result, all of the change in effective potential energy obtained by the shift can be converted into extractable work:
\begin{equation}\label{eq19}
\begin{aligned}
   -W(x,\tau) & = A(x(\tau))-A(x(\tau)-x_f) \;\;\;\; \text{if} \;\; x \geq x_m, &\\
    &=0  \;\;\;\;\;\;\;\;\; \text{if} \;\; x < x_m.
\end{aligned}
\end{equation}

The average extractable work can, therefore,  be obtained as:
\begin{equation}\label{eq20}
-\langle W (\tau) \rangle  =\int_{x_m}^{x_r} dx^{\prime} P_{ss}(x^{\prime},\tau)W(x^{\prime},\tau).
\end{equation}
Here, $P_{ss}(x,\tau)$  denotes the marginal stationary probability distribution of the particle at after relaxation duration $\tau$. At this stage, it is important to mention that particles do not perform any work but are transported during feedback cycles. In the present set-up, no suitable force exists along the feedback direction that defines a nonzero work.  However, we can estimate the amount of the extractable work from the change in effective potential and the related change in obtained information. One way to demonstrate the existence of such an extractable limit is by setting a weak opposing force opposite to the feedback direction \cite{Ali2024jpcb,ali2025ijp}. We use a modified Euler method \cite{xavier1994fortran} to simulate the overdamped two-dimensional Langevin equation inside the confinement with reflecting boundary conditions at the wall. We choose the time step within the range of  $10^{-3}$ to $10^{-4}$ units. We use the Box-Muller algorithm \cite{Box1958ams} to generate Gaussian white noise.  
Unless mentioned otherwise, we use $c = 0.2$, $a = 0.0125$, and $\beta = 1.0$. We calculate the steady state marginal probability distribution ($P_{ss}(x,\tau)$) using a large number ($1\times10^7-3\times10^7$) of trajectories. All numerical integration is performed using Simpson's $1/3$rd rule with a grid size of $10^{-1}$ units. 

\section{Results and Discussion}
\subsection{Steady state 
 marginal probability distribution}
\begin{figure}[!htp]
    \includegraphics[width=0.38\textwidth]{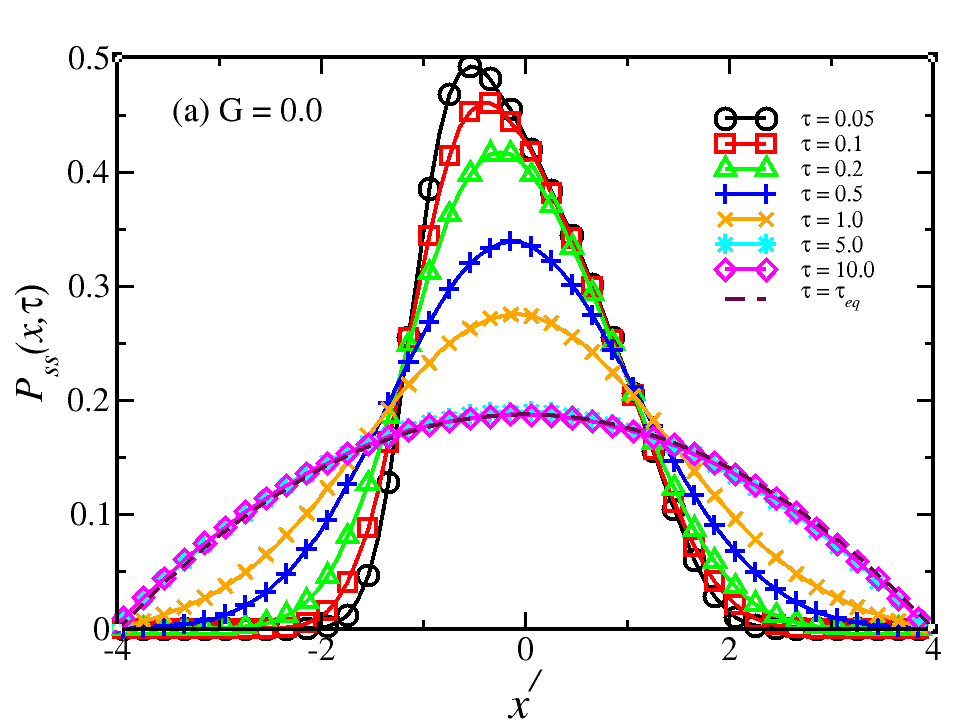}
    \includegraphics[width=0.38\textwidth]{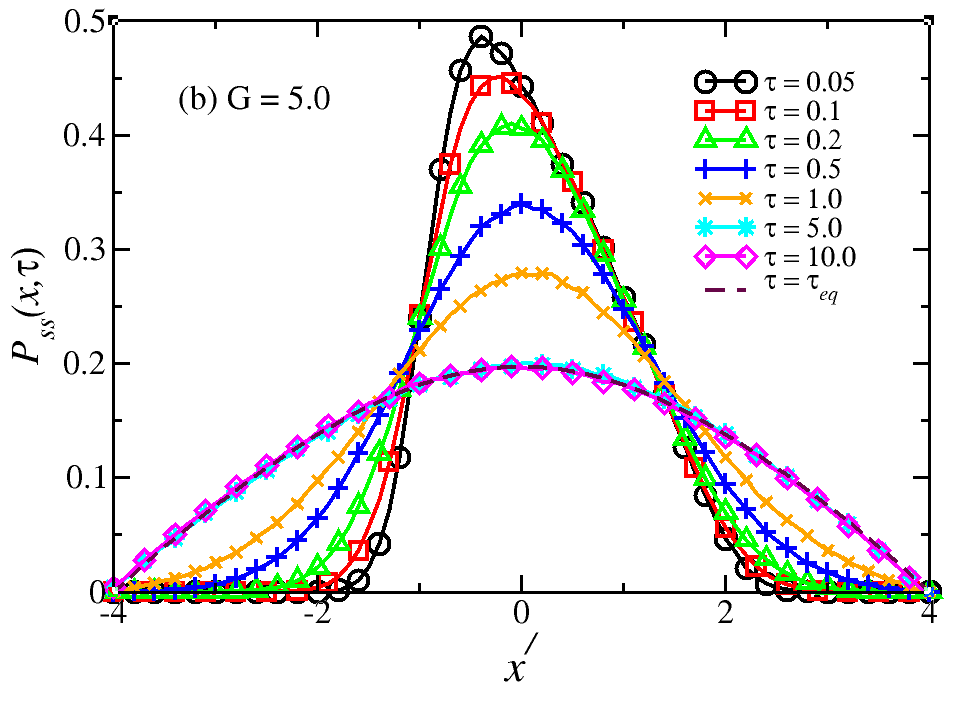}
    \includegraphics[width=0.38\textwidth]{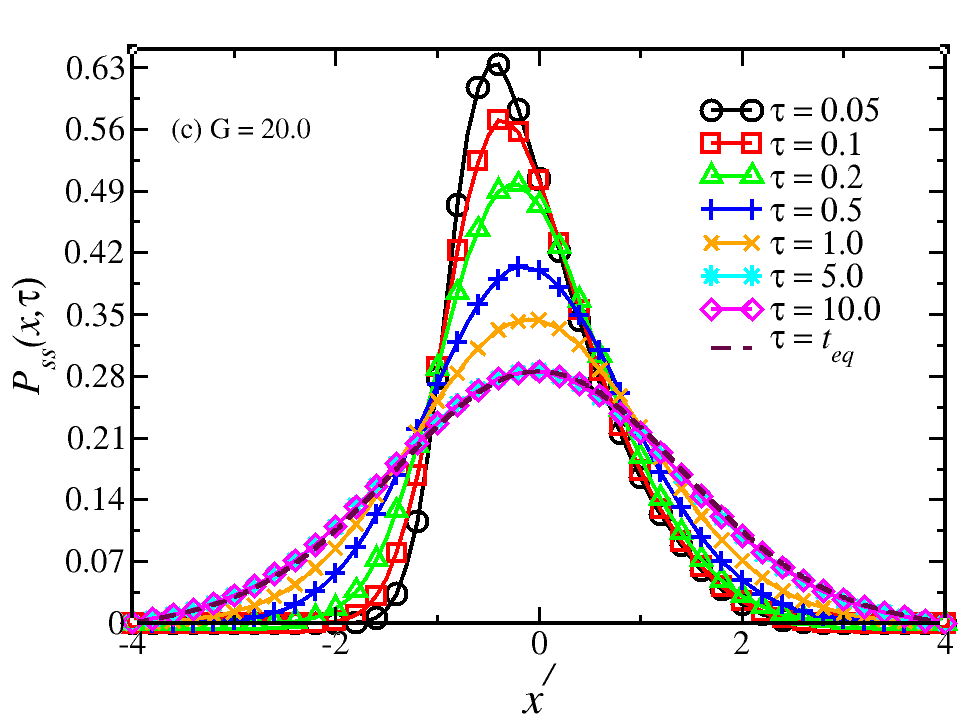}
    \caption{ Time evolution of the steady-state marginal probability distribution function ($P_{ss}(x,\tau)$) for (a) $G = 0.0$, (b) $G = 5.0$ and (c) $G = 20.0$ at different relaxation times ($\tau$). Parameter set chosen: $\beta = 1$, $a = 0.0125$ and $c = 0.2$, for all the cases. All dashed lines (coincide with the results of $\tau=5$) refer to the result in the long time limit ($\tau=\tau_{eq}$) and are obtained from Eq.~\ref{eq10}. }
    \label{f3}
\end{figure}

 Estimating the steady-state marginal probability distribution ($P_{ss}(x,\tau)$) at a cycle time ($\tau$) is crucial to study the temporal evolution of thermodynamic responses of the information engine. Therefore, we begin with the estimation of $P_{ss}(x,\tau)$ for three different strengths of the external bias force ($G$) as shown in Fig.~\ref{f3}. We calculate the same numerically by solving the overdamped Langevin Eq.~(\ref{eq1}) at different relaxation time $(\tau)$.
Fig.~\ref{f3}(a) shows that, in a low $\beta G\omega(x)$ limit and with an increase in $\tau$, the density profile changes in time from an asymmetric (deformed) distribution to a symmetric parabolic one. This implies that the particles are exploring the entire phase space of the confinement once sufficient time (greater than relaxation time $\tau_r$) is provided. On the other hand,  in a high $\beta G\omega(x)$ limit, the strong transverse bias force restricts the particle towards the close vicinity of the lower wall of the confinement. Hence, the shape of the distribution changes from an asymmetric (deformed) distribution to a symmetric one (truncated Gaussian-like) with increasing relaxation time (Fig.~\ref{f3}(c)). In both cases, the system does not get enough time to relax if the $\tau$ is low. Therefore, the distribution spread evolves in time, and the standard deviation $(\sigma)$ approaches a stationary value for a high value of $\tau$ (higher than the relaxation time $(\tau_r)$ of the particle). One can also notice that the relaxation time $(\tau_r)$ remains almost invariant to the extent of entropic dominance ($\tau_r \sim 5)$ units for all three $G$ values under consideration).  Fig.~\ref{f3}(b) depicts the time evolution of $P_{ss}(x,\tau)$ under the influence of moderate energy control ($\beta G = 5$). One can also notice that the relaxation time ($\tau_r$) remains almost invariant to the extent of entropic dominance ($\tau_r \sim 5$ units for all three $G$ values under consideration). \\
 
 In a long time limit, $P_{ss}(x,\tau)$ merges to equilibrium marginal probability distribution function ($P_{eq}(x)$). One can estimate the shape of the distribution $P_{eq}(x)$ theoretically by solving the Smoluchowski equation (Eq.~(\ref{eq14})) in reduced dimension \cite{Ali2022jcp}: 
\begin{equation}\label{eq10}
    \begin{aligned}
         P_{eq}(x)= N\exp\left[{-\beta A(x)}\right],
    \end{aligned}
\end{equation}
  where $N$ is the normalization constant and
   \begin{align*}
     N^{-1} = \frac{1}{\beta G}\sqrt{\frac{\pi }{\beta aG}}\left [ e^{\beta Gc}erf(z)-e^{-\beta Gc}erfi(z)\right],
   \end{align*} 
   with $erf(z)$ ($=\frac{2}{\pi}\int_{0}^{z}e^{-t^2}dt$) and $erfi(z)$ ($=\frac{2}{\pi}\int_{0}^{z}e^{t^2}dt$) 
 are the error function and imaginary error function, respectively, and with $z=\sqrt{\beta Gc}$. Therefore, under the different extent of entropic control, the $P_{eq}(x)$ can be written as:
\begin{equation}\label{20}
\begin{aligned}
 P_{eq}(x) &= \sqrt{\frac{\beta Ga}{\pi}}\exp\left({-\beta Ga}x^2\right),  \;\;\;\; \text{for} \;\; {\beta G} \gg 1,\\
 &= \frac{3}{4}\sqrt{\frac{a}{c^3}}(-ax^2+c),  \;\;\;\; \text{for} \;\; {\beta G} \ll 1.
\end{aligned}
\end{equation}
We have shown the equilibrium results with a brown-colored dashed line in Fig.~\ref{f3}. The observed variation matches well with the numerical simulation data obtained by solving Langevin dynamics over a long time. 
In both entropic and energetic regions, the steady-state marginal probability distribution is asymmetric at low $\tau$ values. Immediately after the measurement, we instantaneously shift the confinement center to $x_f$. Following this feedback step, the particle undergoes relaxation for a duration of $\tau$ with a fixed trap centered at $x_f$. When $\tau$ is low, particles do not have enough time to relax fully. As cycle time increases, the asymmetries decrease.
The extent of asymmetry at low cycle time is higher in the entropy-dominated (low $G$) scenario with respect to a situation of energetic confinement. In this case, particles are exploring more phase space at any instant ($t$). It is worth mentioning that the extent of asymmetry in the marginal probability distribution for a low $\tau$ value depends on the nature of the feedback.
\subsection{Condition for the best work extraction}
Next, we address the strategy to set the best measurement distance ($x_m^*$) and the feedback location ($x_f^*$), that corresponds to the maximum extractable work. Recent studies \cite{Paneru2018prl,Park2016pre,Rafna2023pre,Rafna2024jcp} identify that the combination of $x_m\sim 0.6 \sigma$ and $x_f = 2x_m$ provides the requisites for the best output at equilibrium conditions. Where $\sigma$ denotes the standard deviation of the particle's position. To make the present manuscript self-sufficient, we briefly consult the optimization procedure related to a limiting condition (an energy-dominated situation). In an energy controlled region ($\beta G \gg 1$), the averaged extractable work reads as:
\begin{equation}\label{eq21}
    \begin{aligned}
       -\langle W (\tau)\rangle & = \int_{x_m}^{x_r} dx^{\prime} P_{ss}(x^{\prime},\tau)W(x^{\prime},\tau), &\\
        & = 2aGx_f\int_{x_m}^{x_r} \left(x^{\prime}-\frac{x_f}{2}\right)P_{ss}(x^{\prime},\tau) dx^{\prime}, &\\
        &=2aGx_f\left( \langle x \rangle_r-\frac{x_f}{2}\right)P_r.
    \end{aligned}
\end{equation}
Where $P_{ss}(x,\tau)$ is the steady state marginal probability distribution with relaxation time $\tau$, $P_r=\int_{x_m}^{x_r} dx P_{ss}(x, \tau)$ is the probability when the particle position is larger or equal to $x_m$, and $\langle x \rangle_r$($ =\frac{1}{P_r}\int_{x_m}^{x_r} x P_{ss}(x, \tau) dx$) is the mean position of the particle. The relation implies that the engine produces output work only when  $0<x_f<2\langle x \rangle_r$ at any arbitrary time. At equilibrium ($ \tau \rightarrow \infty$),
\begin{eqnarray}\label{eq22}
   \langle x \rangle_r=\sqrt{\frac{1}{\beta \pi a G}}\frac{\exp{\left({-\beta Gax_m^2}\right)}}{erfc\left(\sqrt{\beta aG}x_m\right)}, \nonumber \\ \text{and}\; \; \;
   P_r=\frac{1}{2}erfc\left(\sqrt{\beta aG}x_m\right),
\end{eqnarray}
where $erfc(z)$ is the complementary error function. Using the value of  $\langle x \rangle_r$ and $P_r$ in Eq.~(\ref{eq21}), 
\begin{eqnarray}\label{eq23}
    -\left \langle W \right \rangle_{eq}&=& x_f Ga \sqrt{\frac{1}{\beta G\pi a}}\exp{\left(-{\beta Gax_m^2}\right)} \nonumber \\
    &-&\frac{Gax_f^2}{2}erf\left(\sqrt{\beta Ga}x_m\right).
\end{eqnarray}
In the spirit of recent studies \cite{Rafna2023pre,Rafna2024jcp}, we obtain the best $x_f^\ast$ and $x_m^\ast$ as follows:
    \begin{equation}\label{eq24a}
      x_f^\ast = \left \langle x \right \rangle_r|_{x_m=x_m^\ast}=
         2x_m^\ast.
\end{equation}
Combining the relations in  Eq.~(\ref{eq23}), we find the dependence of $x_m^\ast$ on other structural parameters as: 
\begin{equation}\label{eq24}
    x_m^\ast = \frac{1}{2} \sqrt{\frac{1}{\beta \pi aG}}\frac{\exp{\left({-\beta Gax_m^{\ast^2}}\right)}}{erfc\left(\sqrt{\beta Ga}x_m^\ast\right)}.
\end{equation}
We solve Eq.~(\ref{eq24})  numerically to get the value of $x_m^\ast \sim 0.6\sigma$, where the standard deviation $\sigma=\sqrt{\frac{1}{2\beta Ga}}$ in this limit. The criteria for best work extraction under an arbitrary strength of $G$ can be obtained numerically following a similar procedure \cite{Park2016pre,Rafna2023pre,Rafna2024jcp}.

\begin{figure}[!h]
    \centering
     \includegraphics[width=0.45\textwidth]{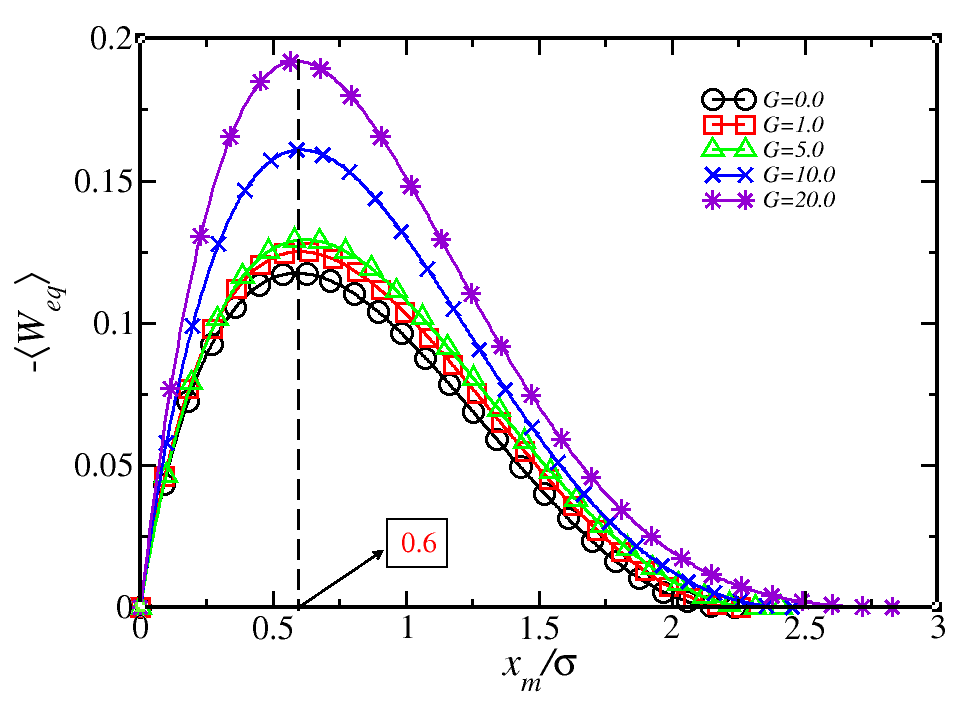}
    \caption{Variation of extractable work with scaled measurement location ${x}_m/\sigma$ under equilibrium conditions for $G = 0.0$ (black colored circular points), $G = 1.0$ (red colored squares), $G = 5.0$ (green colored triangles), $G = 10.0$ (blue colored plus) and $G = 20.0$ (indigo colored stars). Parameter set chosen: $x_f = 2x_m$, $\beta = 1$, $a = 0.0125$ and $c = 0.2$, for all the cases. Lines reflect the theoretical solution derived from Eq.~\ref{eq23}, whereas points represent the numerical simulation obtained from Langevin dynamics simulation.}
    \label{f4}
\end{figure}
 In Fig.~\ref{f4}, we show the variation of extracted work under equilibrium conditions ($\tau \to \infty$) as a function of measurement distance ${x}_m/\sigma$ for different values of $G$. The results show that the maximum amount of work extraction happens at ${x}_m\sim0.6\sigma$, irrespective of the extent of entropic dominance. Here we set $x_f=2x_m$. Thus, in the present study, we fix the measurement distance at $x_m = 0.6\sigma$ and explore a variety of other physical observables with increasing cycle time $\tau$. In the rest of the study, we use a scaled feedback location and scaled measurement distance as  $\tilde{x}_f:={x_f}/{0.6\sigma}$ and  $\tilde{x}_m:={x_m}/{0.6\sigma}$, respectively.
\begin{figure}[!h]
    \centering
    \includegraphics[width=0.40\textwidth]{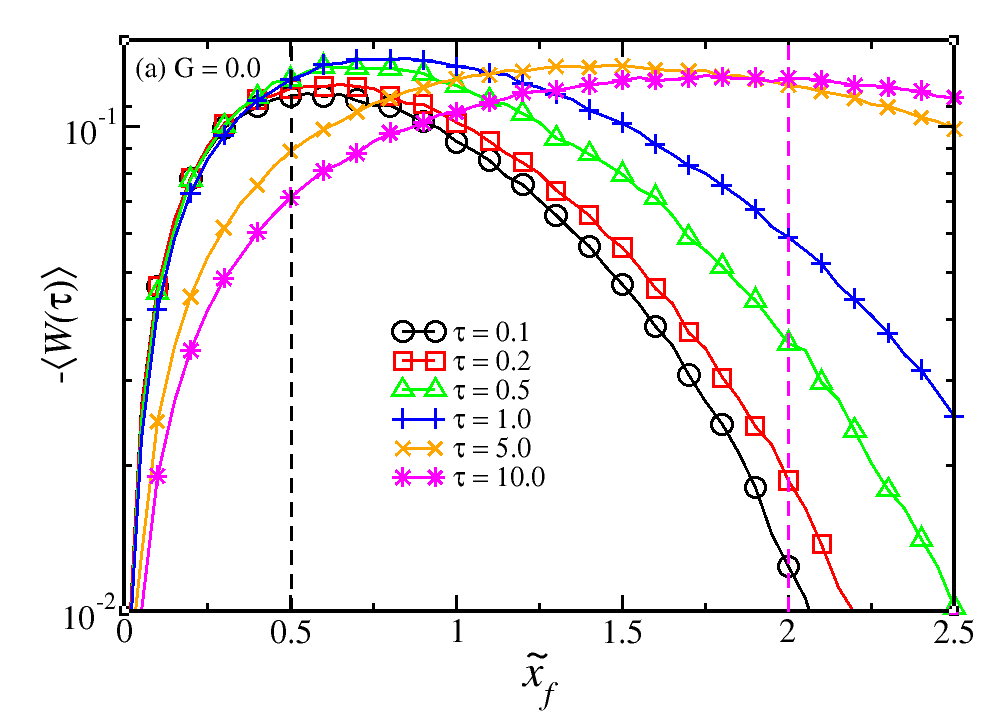}
    \includegraphics[width=0.40\textwidth]{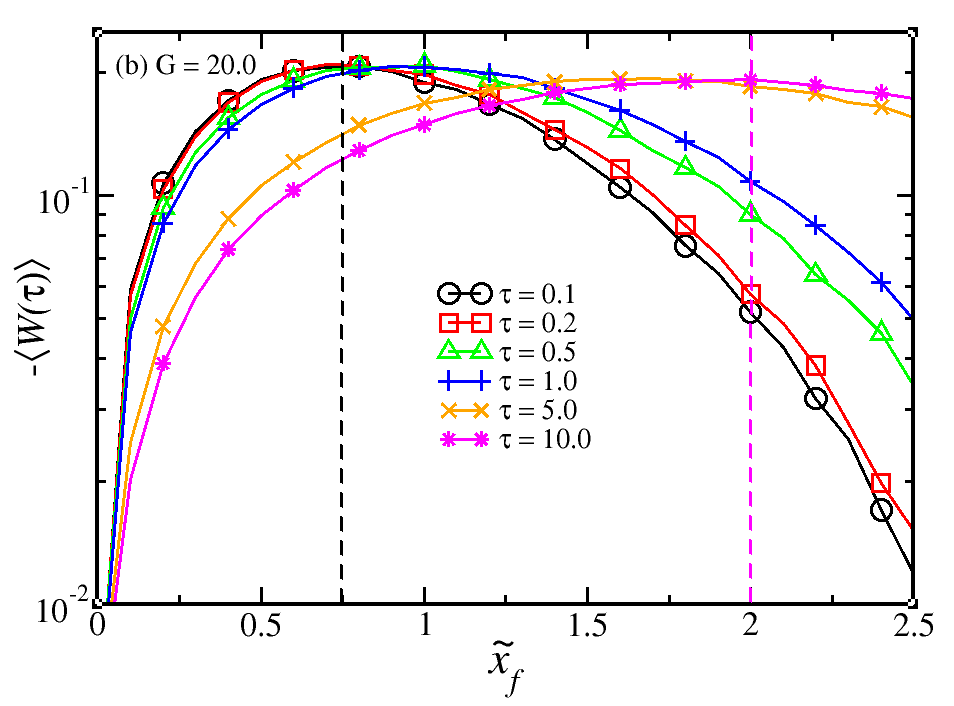}
    \caption{Variation of average extractable work ($-\langle  W (\tau) \rangle$) with scaled feedback site $\tilde{x}_f$ for different cycle period time $\tau$ and for (a) an entropy dominated situation $G = 0.0$ and (b) an energy dominated situation $G = 20.0$. Parameter set chosen: $\beta = 1.0$, $a = 0.0125$ and $c = 0.2$, for all the cases.}
    \label{f5}
\end{figure}

\subsection{Finite time work extraction}
 Proceeding further, we study the variation of the extractable work $(-\langle  W(\tau) \rangle)$ with scaled feedback location ($\tilde{x}_f$) with a fixed measurement distance $x_m = 0.6\sigma$ and at different cycle period time ($\tau$). The results are shown in Fig.~\ref{f5}. The variations depict that the best feedback location $\tilde{x}_f$ at which maximum work can be extracted changes with increasing cycle time ($\tau$). With the present parameter set, if the cycle period time is low (qualitatively $\tau < \tau_r$), the best work extraction is possible with $\tilde{x}_f < 2$. For a fully relaxed state, $(-\langle  W (\tau) \rangle)$ is the maximum for $\tilde{x}_f = 2$, irrespective of the extent of entropic control ($G$). For a low value of $\tilde{x}_f$,  extractable work is higher if the system is not fully relaxed (low $\tau$ value). On the other hand, for a high value of scaled feedback site (high $\tilde{x}_f$) higher work extraction is possible once the system is fully relaxed (high $\tau$). This nontrivial dependency of the extractable work for different cycle times can be explained in terms of the net (available) information acquired during the feedback. We will discuss this issue in the next subsection. Figures also show that the average extractable work in a fully relaxed state is higher for an energy-dominated situation. To explain this observation, we recall that the average extractable work per cycle depends on the difference between the total information and the unavailable information lost due to the relaxation process \cite{Ali2022jcp}. One can extract as much information as possible by instantaneously shifting the confinement and corresponding potential centres.
Now, unavailable information arises because of the relaxation process after the feedback. With increasing $G$, the unavailable information $(\left \langle  I_u \right \rangle)$ decreases, as discussed in \cite{Ali2022jcp}. Therefore, the available information that can be converted to an extractable work is higher with high values of $G$.

\subsection{Finite time power extraction}
Next, we calculate the average extracted power ($ \left \langle Po (\tau)\right \rangle$). The power at time $\tau$  can be defined as:
\begin{equation}\label{eq26}
    \left \langle  Po (\tau) \right \rangle=-\frac{\left \langle  W (\tau) \right \rangle}{\tau}.
\end{equation}
We plot the average extractable power as a function of $\tilde{x}_f$ for different $\tau$  as shown in Fig.~\ref{f6}. The observations are following:
\begin{figure}[!h]
    \centering
    \includegraphics[width=0.40\textwidth]{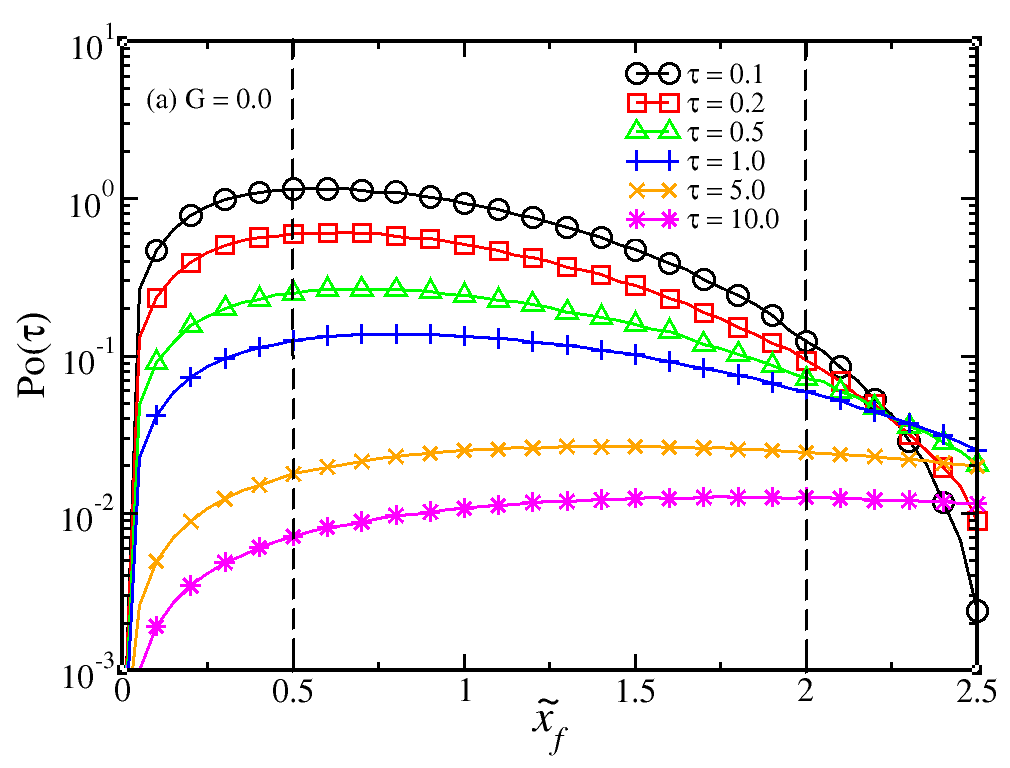}
    \includegraphics[width=0.40\textwidth]{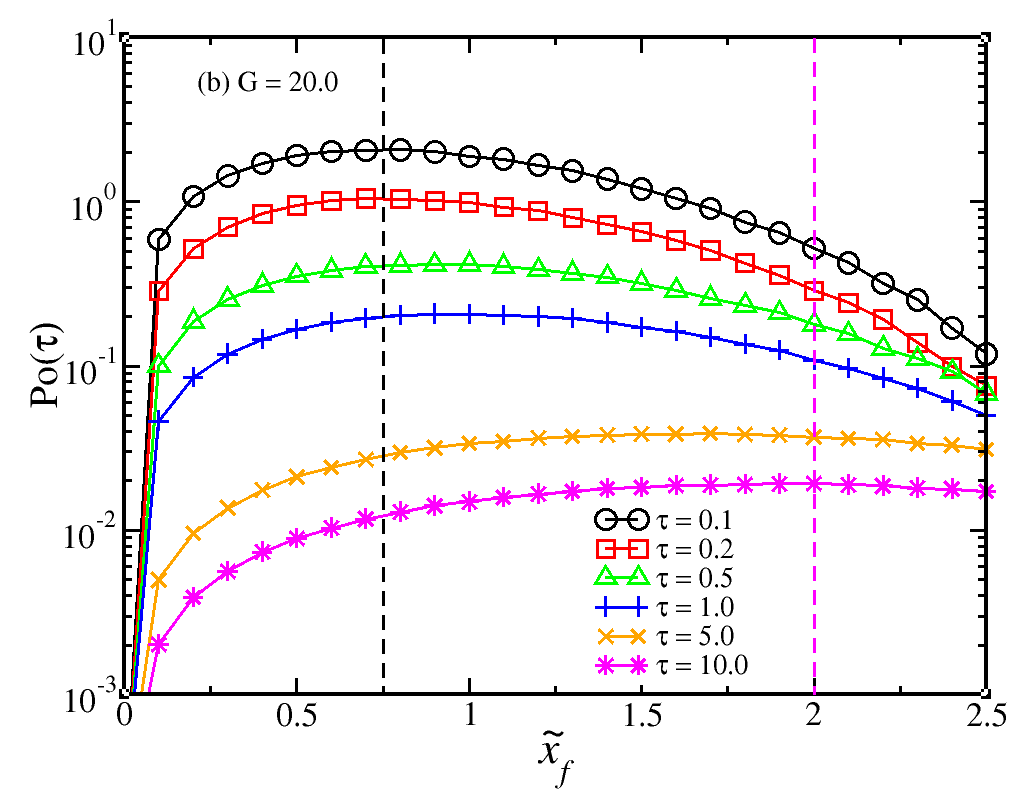}
   
  \caption{Variation of average extracted power ($\langle  Po (\tau) \rangle$) with scaled feedback location ($\tilde{x}_f$) in different cycle time $\tau$. (a) Entropy-dominated $(G = 0.0)$, and (b) Energy-dominated $(G = 20.0)$ scenarios Parameter set chosen: $\beta = 1$, $a = 0.0125$ and $c = 0.2$, for all the cases.}
    \label{f6}
\end{figure}
First, within the range of cycle times under consideration, the magnitude of maximum average obtainable power $(\langle Po (\tau) \rangle)$ decreases with increasing $\tau$. The result is obvious, as the amount of extractable work does not increase rapidly while increasing $(\tau)$. When the system completely relaxes $(\tau \ge \tau_r)$, the amount of extractable work saturates to the maximum possible value. Subsequently, the power decreases with a further increase in $\tau$ (for $\tau \ge \tau_r$).  Second,  when the cycle time is less than the characteristic relaxation time $(\tau<\tau_r)$,  the obtainable power reaches a maximum for a scaled feedback location $\tilde{x}_f<2$, irrespective of the entropic control.  Third, upon increasing the feedback cycle time, the position of the maximum shifted towards a higher value of the feedback site and reached $\tilde{x}_f=2$ for a fully relaxed state. Finally, one may observe that the magnitude of maximum power $(\langle  Po \rangle_{max})$ is lower in an entropy-controlled process (in comparison to an energy-dominated one). Such a lowering of the power can be explained by considering the following argument. The maximum extractable work is less, and the relaxation time scale is almost similar in the entropy-dominated limit compared to the energy-controlled situation. Therefore, the maximum value of  $\left \langle  Po (\tau) \right \rangle$ is less for a pure entropic GBIE. 


\begin{figure}[!h]
    \centering
    \includegraphics[width=0.4\textwidth]{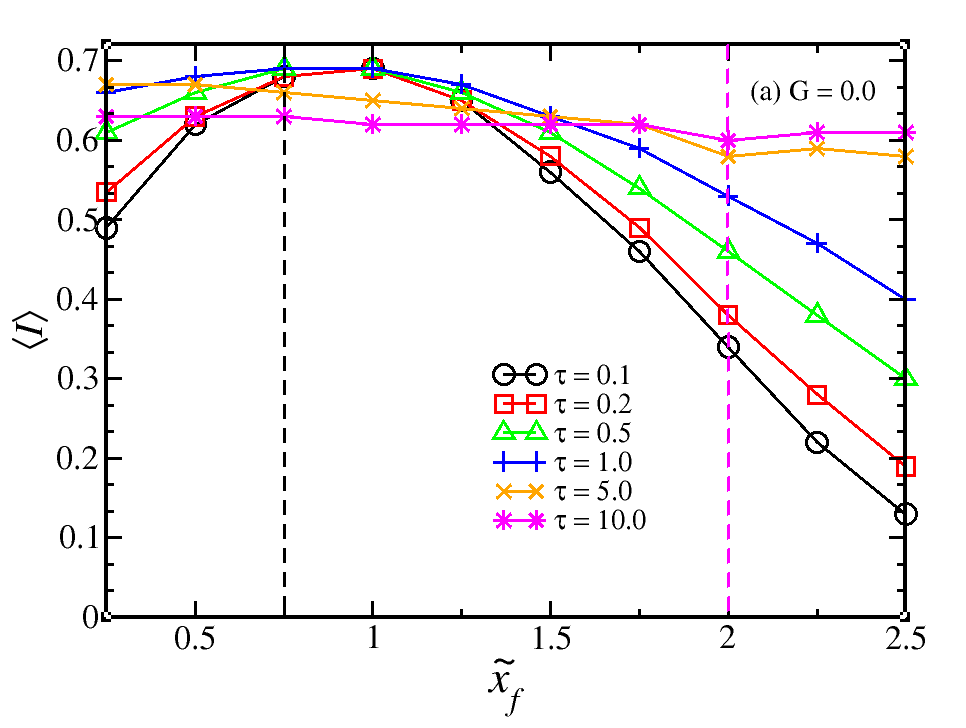}
   \includegraphics[width=0.4\textwidth]{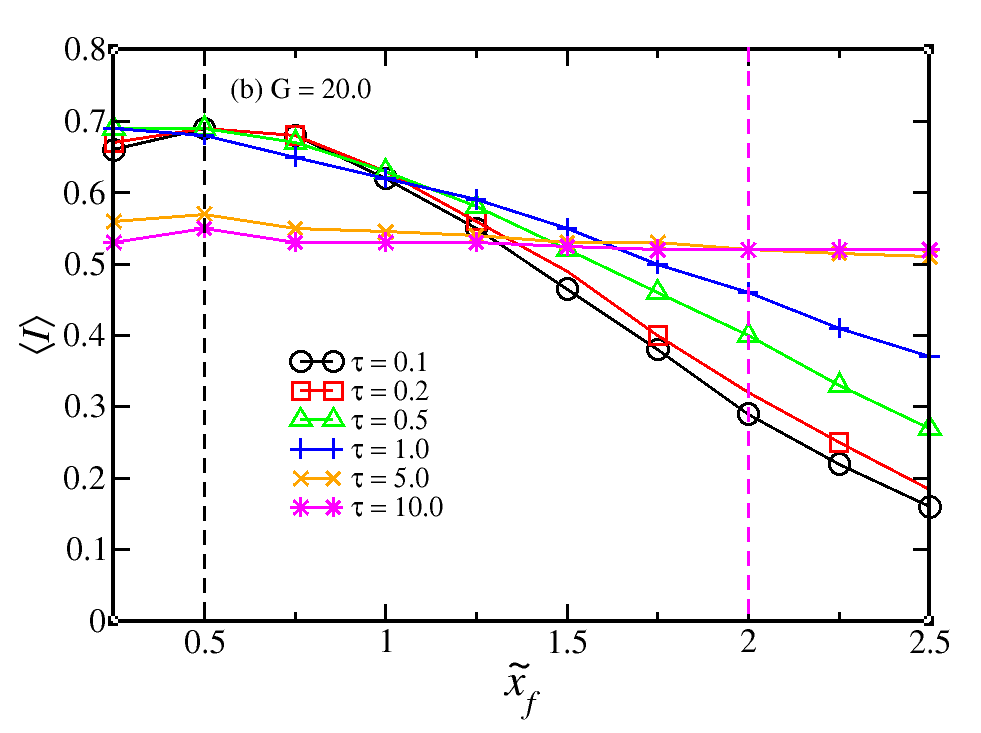}
    \caption{The variation of the average total information $(\langle I\rangle )$  with scaled feedback location $\tilde{x}_f$ in different cycle times $\tau$ for (a) $G = 0.0$ and (b) $G = 20$. Parameter set chosen: $\tilde{x}_m = 1$, $\beta = 1$, $a = 0.0125$ and $c = 0.2$ for all cases.}
   \label{f7}
\end{figure}

\subsection{Information and efficiency}

Finally, we estimate the time evolution of the efficiency of a GBIE under different entropic dominance. The efficiency is defined as $\eta (\tau) = {-\left \langle  W (\tau) \right \rangle}/{\left \langle I \right \rangle}$. Where $\left \langle I \right \rangle$ is the information obtained during the measurement \cite{Bauer2012jphysa} step, averaged over the number of cycles. Here, we consider an error-free (almost) measurement process. Under such constraint, one can estimate the information grossed during the measurement as the average Shannon entropy of the particle.
The present feedback scheme has two discrete measurement outcomes: either a particle is found on the right-hand side of $x_m$ with probability $P_r$ or on the left side of $x_m$ with probability $1-P_r$. Thus, one can measure the acquired information as $\left \langle I \right \rangle$ = $-P_r \ln P_r-(1-P_r) \ln(1-P_r)$ \cite{Paneru2018pre}.\\

The variation of information with $(\langle I\rangle)$ with scaled feedback location $\tilde{x}_f$ in different cycle time $\tau$ for $G = 0.0$ and $G = 20$ is shown in Fig.~\ref{f7}. When the feedback location is high $\tilde{x}_f > 2$ the acquired information is higher for a fully relaxed state. On the contrary, if the feedback location is set at a very low value $\tilde{x}_f < 2$, acquired information decreases with an increase in cycle time. This variation in acquired information explains the nontrivial cycle time dependency of the extractable work. Also, one can notice that the non-monotonicity of $(\langle I\rangle)$ at the small cycle time region is more pronounced in the entropy-dominated situation. \\

\begin{figure}[!h]
    \centering
   \includegraphics[width=0.4\textwidth]{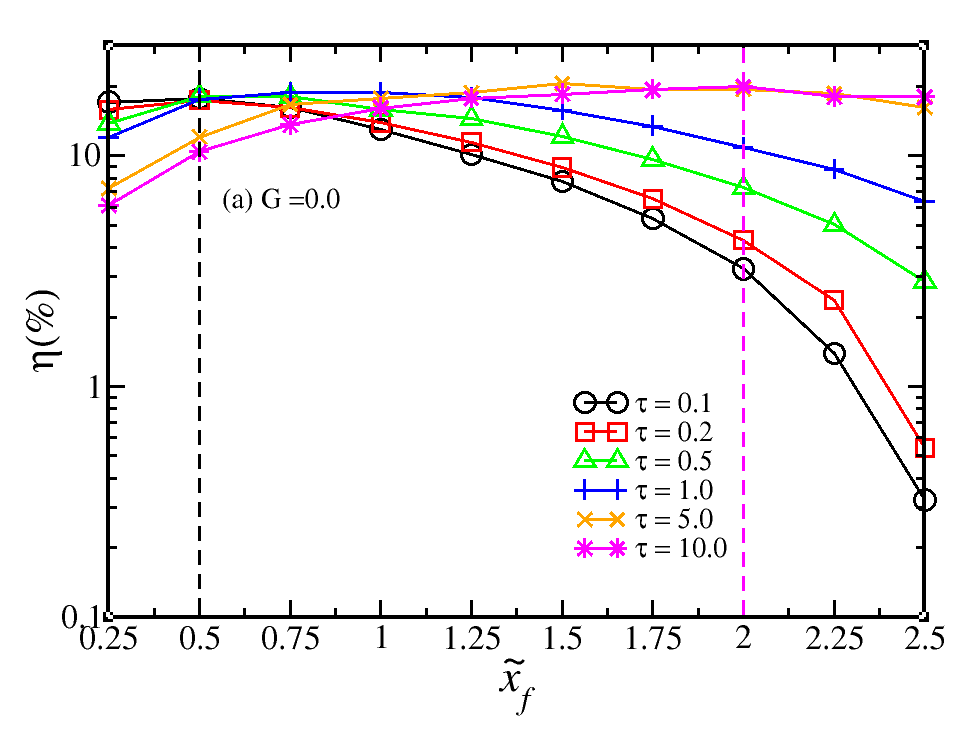}
   \includegraphics[width=0.4\textwidth]{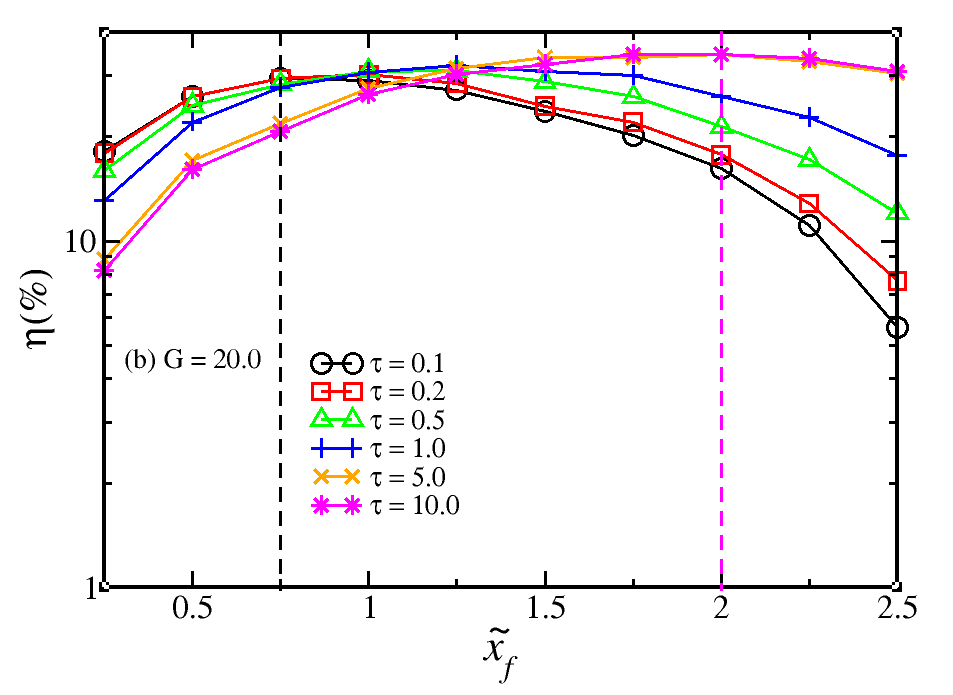}
    \caption{ The variation of the efficiency percentage ($\eta \times 100$)  with scaled feedback location $\tilde{x}_f$ in different cycle time $\tau$ for (a) $G = 0.0$ and (b) $G = 20$. Parameter set chosen: $\tilde{x}_m = 1$, $\beta = 1$, $a = 0.0125$ and $c = 0.2$ for all cases.}
    \label{f8}
\end{figure}

Finally, Fig.~\ref{f8} depicts the variation of the engine's efficiency ($\eta$) as a function of the scaled feedback location $\tilde{x}_f$ at different cycle times $\tau$ (in a percentage scale). For a given cycle time $\tau $, when the scaled feedback location is low,  the efficiency increases with an increase in $\tilde{x}_f$. The efficiency shows a maximum at some intermediate $\tilde{x}_f$ $(< 2.0)$ and then decreases. When the feedback site is set at a lower value, a low cycle time $(\tau < \tau_r)$ results in a higher efficiency. However, for $\tilde{x}_f>2$, the efficiency increases with increasing $\tau$ and saturates to a maximum at a high time limit $\tau \ge \tau_r$.
In a long time limit $\tau \ge \tau_r$, the maximum of efficiency is thus obtained at $\tilde{x}_f=2.0$ or ${x}_f=2x_m$, irrespective of the extent of the entropic control. Finally, we find that, for a fully relaxed state, the maximum efficiency $\eta_{max}$ at an entropy-dominated $(G \rightarrow 0)$ situation is $\simeq20.0\%$, whereas under energetic control $(G = 20.0)$,  $\simeq35\%$. A higher magnitude of the maximum extractable work and a lower amount of the acquired information in case of an energy-dominated situation (at a fully relaxed state) can explain this observation.\\

\section{Conclusion}

In conclusion, we study the temporal evolution of a Geometric Brownian Information Engine (GBIE) operating from non-equilibrium steady-state conditions to a fully relaxed state with a finite cycle duration. Engine cycles consist of three stages: measurement, feedback, and relaxation.  We optimise thermodynamic observables such as extractable work, power, and efficiency using an error-free feedback control mechanism in both entropic and energetic dominance scenarios. The steady-state marginal probability distribution changes from asymmetric to symmetric with an increase in the finite cycle time $\tau$.
\par
We find that the best feedback location to maximize extractable work changes with varying feeback cycle times in the presence of a given (optimal) measurement distance ($x_m \sim 0.61\sigma$). For shorter cycle periods, the ideal feedback site is less than 2 times $x_f$. The optimal feedback position approaches double the measurement distance as $\tau$ increases and the system gets more time to be relaxed. We also observe that the average efficiency of the GBIE increases with increasing feedback location up to an intermediate value and then decreases. In the long time limit, the maximum efficiency is achieved at a feedback location of 2, regardless of the extent of entropic control. Because of the higher loss of information during relaxation, the extractable work and the efficiency in the fully relaxed state decrease with the increasing dominance of the entropic control.

Further, we look at the GBIE's capacity to extract power. There is a decrease in the maximum average power as the cycle time increases. In the case of cycle times that are shorter than the characteristic relaxation time, the maximum power is achieved when the scaled feedback locations are less than 2. Conversely, for cycle times that exceed the characteristic relaxation time, the maximum power is attained when the scaled feedback location is precisely at 2. In the case of an entropy-controlled device, the amount of extractable work is lower. With the chosen control parameters, an energy-dominated Information engine can generate almost ten times more power than the other extreme. Our study sheds light on the time-dependent behaviour of a Geometric Brownian Information Engine and emphasizes the need to optimize system parameters to maximize work extraction, power, and efficiency. These discoveries can help with the design and optimization of future nanoscale devices or artificial biological motors by advancing our understanding of information engines that operate in non-equilibrium situations.

\begin{acknowledgments}
  RR acknowledges  DST INSPIRE [DST/INSPIRE/03/2021/002138] and IIT Tirupati for fellowship. DM thanks SERB (Project No. ECR/2018/002830/CS), Department of Science and Technology, Government of India, for financial support and IIT Tirupati for the computational facility. SYA acknowledges IIT Tirupati and IIT Bombay for fellowship.
\end{acknowledgments}
\section*{Data Availability}
The data that support the findings of this study are available within the article.
\section*{Conflict of interest}
The authors have no conflicts to disclose.
\bibliography{bibfile}

\providecommand{\noopsort}[1]{}\providecommand{\singleletter}[1]{#1}%
\begin{thebibliography}{106}%
\makeatletter
\providecommand \@ifxundefined [1]{%
 \@ifx{#1\undefined}
}%
\providecommand \@ifnum [1]{%
 \ifnum #1\expandafter \@firstoftwo
 \else \expandafter \@secondoftwo
 \fi
}%
\providecommand \@ifx [1]{%
 \ifx #1\expandafter \@firstoftwo
 \else \expandafter \@secondoftwo
 \fi
}%
\providecommand \natexlab [1]{#1}%
\providecommand \enquote  [1]{``#1''}%
\providecommand \bibnamefont  [1]{#1}%
\providecommand \bibfnamefont [1]{#1}%
\providecommand \citenamefont [1]{#1}%
\providecommand \href@noop [0]{\@secondoftwo}%
\providecommand \href [0]{\begingroup \@sanitize@url \@href}%
\providecommand \@href[1]{\@@startlink{#1}\@@href}%
\providecommand \@@href[1]{\endgroup#1\@@endlink}%
\providecommand \@sanitize@url [0]{\catcode `\\12\catcode `\$12\catcode
  `\&12\catcode `\#12\catcode `\^12\catcode `\_12\catcode `\%12\relax}%
\providecommand \@@startlink[1]{}%
\providecommand \@@endlink[0]{}%
\providecommand \url  [0]{\begingroup\@sanitize@url \@url }%
\providecommand \@url [1]{\endgroup\@href {#1}{\urlprefix }}%
\providecommand \urlprefix  [0]{URL }%
\providecommand \Eprint [0]{\href }%
\providecommand \doibase [0]{https://doi.org/}%
\providecommand \selectlanguage [0]{\@gobble}%
\providecommand \bibinfo  [0]{\@secondoftwo}%
\providecommand \bibfield  [0]{\@secondoftwo}%
\providecommand \translation [1]{[#1]}%
\providecommand \BibitemOpen [0]{}%
\providecommand \bibitemStop [0]{}%
\providecommand \bibitemNoStop [0]{.\EOS\space}%
\providecommand \EOS [0]{\spacefactor3000\relax}%
\providecommand \BibitemShut  [1]{\csname bibitem#1\endcsname}%
\let\auto@bib@innerbib\@empty
\bibitem [{\citenamefont {Parrondo}\ \emph {et~al.}(2015)\citenamefont
  {Parrondo}, \citenamefont {Horowitz},\ and\ \citenamefont
  {Sagawa}}]{parrondo2015nat}%
  \BibitemOpen
  \bibfield  {author} {\bibinfo {author} {\bibfnamefont {J.~M.}\ \bibnamefont
  {Parrondo}}, \bibinfo {author} {\bibfnamefont {J.~M.}\ \bibnamefont
  {Horowitz}},\ and\ \bibinfo {author} {\bibfnamefont {T.}~\bibnamefont
  {Sagawa}},\ }\bibfield  {title} {\bibinfo {title} {Thermodynamics of
  information},\ }\href {https://doi.org/https://doi.org/10.1038/nphys3230}
  {\bibfield  {journal} {\bibinfo  {journal} {Nature Phys}\ }\textbf {\bibinfo
  {volume} {11}},\ \bibinfo {pages} {131} (\bibinfo {year} {2015})}\BibitemShut
  {NoStop}%
\bibitem [{\citenamefont {Linke}\ and\ \citenamefont
  {Parrondo}(2021)}]{Linke2021pnas}%
  \BibitemOpen
  \bibfield  {author} {\bibinfo {author} {\bibfnamefont {H.}~\bibnamefont
  {Linke}}\ and\ \bibinfo {author} {\bibfnamefont {J.~M.~R.}\ \bibnamefont
  {Parrondo}},\ }\bibfield  {title} {\bibinfo {title} {Tuning up maxwell’s
  demon},\ }\href {https://doi.org/10.1073/pnas.2108218118} {\bibfield
  {journal} {\bibinfo  {journal} {Proc. Natl. Acad. Sci. U. S. A}\ }\textbf
  {\bibinfo {volume} {118}},\ \bibinfo {pages} {e2108218118} (\bibinfo {year}
  {2021})}\BibitemShut {NoStop}%
\bibitem [{\citenamefont {Horowitz}\ \emph {et~al.}(2013)\citenamefont
  {Horowitz}, \citenamefont {Sagawa},\ and\ \citenamefont
  {Parrondo}}]{Horowitz2013prl}%
  \BibitemOpen
  \bibfield  {author} {\bibinfo {author} {\bibfnamefont {J.~M.}\ \bibnamefont
  {Horowitz}}, \bibinfo {author} {\bibfnamefont {T.}~\bibnamefont {Sagawa}},\
  and\ \bibinfo {author} {\bibfnamefont {J.~M.~R.}\ \bibnamefont {Parrondo}},\
  }\bibfield  {title} {\bibinfo {title} {Imitating chemical motors with optimal
  information motors},\ }\href {https://doi.org/10.1103/PhysRevLett.111.010602}
  {\bibfield  {journal} {\bibinfo  {journal} {Phys. Rev. Lett.}\ }\textbf
  {\bibinfo {volume} {111}},\ \bibinfo {pages} {010602} (\bibinfo {year}
  {2013})}\BibitemShut {NoStop}%
\bibitem [{\citenamefont {Ashida}\ \emph {et~al.}(2014)\citenamefont {Ashida},
  \citenamefont {Funo}, \citenamefont {Murashita},\ and\ \citenamefont
  {Ueda}}]{Ashida2014pre}%
  \BibitemOpen
  \bibfield  {author} {\bibinfo {author} {\bibfnamefont {Y.}~\bibnamefont
  {Ashida}}, \bibinfo {author} {\bibfnamefont {K.}~\bibnamefont {Funo}},
  \bibinfo {author} {\bibfnamefont {Y.}~\bibnamefont {Murashita}},\ and\
  \bibinfo {author} {\bibfnamefont {M.}~\bibnamefont {Ueda}},\ }\bibfield
  {title} {\bibinfo {title} {General achievable bound of extractable work under
  feedback control},\ }\href {https://doi.org/10.1103/PhysRevE.90.052125}
  {\bibfield  {journal} {\bibinfo  {journal} {Phys. Rev. E}\ }\textbf {\bibinfo
  {volume} {90}},\ \bibinfo {pages} {052125} (\bibinfo {year}
  {2014})}\BibitemShut {NoStop}%
\bibitem [{\citenamefont {B{\'e}rut}\ \emph {et~al.}(2012)\citenamefont
  {B{\'e}rut}, \citenamefont {Arakelyan}, \citenamefont {Petrosyan},
  \citenamefont {Ciliberto}, \citenamefont {Dillenschneider},\ and\
  \citenamefont {Lutz}}]{Berut2012nat}%
  \BibitemOpen
  \bibfield  {author} {\bibinfo {author} {\bibfnamefont {A.}~\bibnamefont
  {B{\'e}rut}}, \bibinfo {author} {\bibfnamefont {A.}~\bibnamefont
  {Arakelyan}}, \bibinfo {author} {\bibfnamefont {A.}~\bibnamefont
  {Petrosyan}}, \bibinfo {author} {\bibfnamefont {S.}~\bibnamefont
  {Ciliberto}}, \bibinfo {author} {\bibfnamefont {R.}~\bibnamefont
  {Dillenschneider}},\ and\ \bibinfo {author} {\bibfnamefont {E.}~\bibnamefont
  {Lutz}},\ }\bibfield  {title} {\bibinfo {title} {Experimental verification of
  {L}andauer's principle linking information and thermodynamics},\ }\href
  {https://doi.org/10.1038/nature10872} {\bibfield  {journal} {\bibinfo
  {journal} {Nature}\ }\textbf {\bibinfo {volume} {483}},\ \bibinfo {pages}
  {187} (\bibinfo {year} {2012})}\BibitemShut {NoStop}%
\bibitem [{\citenamefont {Paneru}\ \emph
  {et~al.}(2018{\natexlab{a}})\citenamefont {Paneru}, \citenamefont {Lee},
  \citenamefont {Tlusty},\ and\ \citenamefont {Pak}}]{Paneru2018prl}%
  \BibitemOpen
  \bibfield  {author} {\bibinfo {author} {\bibfnamefont {G.}~\bibnamefont
  {Paneru}}, \bibinfo {author} {\bibfnamefont {D.~Y.}\ \bibnamefont {Lee}},
  \bibinfo {author} {\bibfnamefont {T.}~\bibnamefont {Tlusty}},\ and\ \bibinfo
  {author} {\bibfnamefont {H.~K.}\ \bibnamefont {Pak}},\ }\bibfield  {title}
  {\bibinfo {title} {Lossless {B}rownian {I}nformation {E}ngine},\ }\href
  {https://doi.org/10.1103/PhysRevLett.120.020601} {\bibfield  {journal}
  {\bibinfo  {journal} {Phys. Rev. Lett.}\ }\textbf {\bibinfo {volume} {120}},\
  \bibinfo {pages} {020601} (\bibinfo {year} {2018}{\natexlab{a}})}\BibitemShut
  {NoStop}%
\bibitem [{\citenamefont {Maxwell}(1871)}]{maxwell1871theory}%
  \BibitemOpen
  \bibfield  {author} {\bibinfo {author} {\bibfnamefont {J.~C.}\ \bibnamefont
  {Maxwell}},\ }\bibfield  {title} {\bibinfo {title} {Theory of heat
  longmans},\ }\href@noop {} {\bibfield  {journal} {\bibinfo  {journal} {Green
  and Co, London}\ } (\bibinfo {year} {1871})}\BibitemShut {NoStop}%
\bibitem [{\citenamefont {Leff}\ and\ \citenamefont
  {Rex}(2002)}]{Rex2003maxwell}%
  \BibitemOpen
  \bibfield  {author} {\bibinfo {author} {\bibfnamefont {H.}~\bibnamefont
  {Leff}}\ and\ \bibinfo {author} {\bibfnamefont {A.~F.}\ \bibnamefont {Rex}},\
  }\href@noop {} {\emph {\bibinfo {title} {Maxwell's Demon 2 Entropy, Classical
  and Quantum Information, Computing}}}\ (\bibinfo  {publisher} {CRC Press},\
  \bibinfo {year} {2002})\BibitemShut {NoStop}%
\bibitem [{\citenamefont {Szilard}(1929)}]{Szilard1929zphys}%
  \BibitemOpen
  \bibfield  {author} {\bibinfo {author} {\bibfnamefont {L.}~\bibnamefont
  {Szilard}},\ }\bibfield  {title} {\bibinfo {title} {{\"u}ber die
  {E}ntropieverminderung in einem thermodynamischen {S}ystem bei {E}ingriffen
  intelligenter {W}esen},\ }\href {https://doi.org/10.1007/BF01341281}
  {\bibfield  {journal} {\bibinfo  {journal} {Z. Phys}\ }\textbf {\bibinfo
  {volume} {53}},\ \bibinfo {pages} {840} (\bibinfo {year} {1929})}\BibitemShut
  {NoStop}%
\bibitem [{\citenamefont {Brillouin}(1951)}]{Brillouin1951jcp}%
  \BibitemOpen
  \bibfield  {author} {\bibinfo {author} {\bibfnamefont {L.}~\bibnamefont
  {Brillouin}},\ }\bibfield  {title} {\bibinfo {title} {Maxwell's {D}emon
  {C}annot {O}perate: {I}nformation and {E}ntropy. {I}},\ }\href
  {https://doi.org/10.1063/1.1699951} {\bibfield  {journal} {\bibinfo
  {journal} {J. Appl. Phys.}\ }\textbf {\bibinfo {volume} {22}},\ \bibinfo
  {pages} {334} (\bibinfo {year} {1951})}\BibitemShut {NoStop}%
\bibitem [{\citenamefont {Landauer}(1961)}]{Landaueribm1961}%
  \BibitemOpen
  \bibfield  {author} {\bibinfo {author} {\bibfnamefont {R.}~\bibnamefont
  {Landauer}},\ }\bibfield  {title} {\bibinfo {title} {Irreversibility and
  {H}eat {G}eneration in the {C}omputing {P}rocess},\ }\href
  {https://doi.org/10.1147/rd.53.0183} {\bibfield  {journal} {\bibinfo
  {journal} {IBM J. Res. Dev}\ }\textbf {\bibinfo {volume} {5}},\ \bibinfo
  {pages} {183} (\bibinfo {year} {1961})}\BibitemShut {NoStop}%
\bibitem [{\citenamefont {Bennett}(1982)}]{Bennett1982intjtphys}%
  \BibitemOpen
  \bibfield  {author} {\bibinfo {author} {\bibfnamefont {C.~H.}\ \bibnamefont
  {Bennett}},\ }\bibfield  {title} {\bibinfo {title} {The thermodynamics of
  computation—a review},\ }\href {https://doi.org/10.1007/BF02084158}
  {\bibfield  {journal} {\bibinfo  {journal} {Int. J. Theor. Phys}\ }\textbf
  {\bibinfo {volume} {21}},\ \bibinfo {pages} {905} (\bibinfo {year}
  {1982})}\BibitemShut {NoStop}%
\bibitem [{\citenamefont {Sagawa}\ and\ \citenamefont
  {Ueda}(2008)}]{Sagwa2008prl}%
  \BibitemOpen
  \bibfield  {author} {\bibinfo {author} {\bibfnamefont {T.}~\bibnamefont
  {Sagawa}}\ and\ \bibinfo {author} {\bibfnamefont {M.}~\bibnamefont {Ueda}},\
  }\bibfield  {title} {\bibinfo {title} {Second {L}aw of {T}hermodynamics with
  {D}iscrete {Q}uantum {F}eedback {C}ontrol},\ }\href
  {https://doi.org/10.1103/PhysRevLett.100.080403} {\bibfield  {journal}
  {\bibinfo  {journal} {Phys. Rev. Lett.}\ }\textbf {\bibinfo {volume} {100}},\
  \bibinfo {pages} {080403} (\bibinfo {year} {2008})}\BibitemShut {NoStop}%
\bibitem [{\citenamefont {Sagawa}\ and\ \citenamefont
  {Ueda}(2009)}]{Sagawa2009prl}%
  \BibitemOpen
  \bibfield  {author} {\bibinfo {author} {\bibfnamefont {T.}~\bibnamefont
  {Sagawa}}\ and\ \bibinfo {author} {\bibfnamefont {M.}~\bibnamefont {Ueda}},\
  }\bibfield  {title} {\bibinfo {title} {Minimal {E}nergy {C}ost for
  {T}hermodynamic {I}nformation {P}rocessing: {M}easurement and {I}nformation
  {E}rasure},\ }\href {https://doi.org/10.1103/PhysRevLett.102.250602}
  {\bibfield  {journal} {\bibinfo  {journal} {Phys. Rev. Lett.}\ }\textbf
  {\bibinfo {volume} {102}},\ \bibinfo {pages} {250602} (\bibinfo {year}
  {2009})}\BibitemShut {NoStop}%
\bibitem [{\citenamefont {Sagawa}\ and\ \citenamefont
  {Ueda}(2010)}]{Sagawa2010prl}%
  \BibitemOpen
  \bibfield  {author} {\bibinfo {author} {\bibfnamefont {T.}~\bibnamefont
  {Sagawa}}\ and\ \bibinfo {author} {\bibfnamefont {M.}~\bibnamefont {Ueda}},\
  }\bibfield  {title} {\bibinfo {title} {Generalized {J}arzynski {E}quality
  under {N}onequilibrium {F}eedback {C}ontrol},\ }\href
  {https://doi.org/10.1103/PhysRevLett.104.090602} {\bibfield  {journal}
  {\bibinfo  {journal} {Phys. Rev. Lett.}\ }\textbf {\bibinfo {volume} {104}},\
  \bibinfo {pages} {090602} (\bibinfo {year} {2010})}\BibitemShut {NoStop}%
\bibitem [{\citenamefont {Jarzynski}(1997)}]{jarzynski1997prl}%
  \BibitemOpen
  \bibfield  {author} {\bibinfo {author} {\bibfnamefont {C.}~\bibnamefont
  {Jarzynski}},\ }\bibfield  {title} {\bibinfo {title} {Nonequilibrium
  {E}quality for {F}ree {E}nergy {D}ifferences},\ }\href
  {https://doi.org/10.1103/PhysRevLett.78.2690} {\bibfield  {journal} {\bibinfo
   {journal} {Phys. Rev. Lett.}\ }\textbf {\bibinfo {volume} {78}},\ \bibinfo
  {pages} {2690} (\bibinfo {year} {1997})}\BibitemShut {NoStop}%
\bibitem [{\citenamefont {Sekimoto}(2010)}]{sekimoto2010stochastic}%
  \BibitemOpen
  \bibfield  {author} {\bibinfo {author} {\bibfnamefont {K.}~\bibnamefont
  {Sekimoto}},\ }\href@noop {} {\bibinfo {title} {Stochastic energetics}}
  (\bibinfo {year} {2010})\BibitemShut {NoStop}%
\bibitem [{\citenamefont {Seifert}(2012)}]{seifert2012rpp}%
  \BibitemOpen
  \bibfield  {author} {\bibinfo {author} {\bibfnamefont {U.}~\bibnamefont
  {Seifert}},\ }\bibfield  {title} {\bibinfo {title} {Stochastic
  thermodynamics, fluctuation theorems and molecular machines},\ }\href
  {https://doi.org/10.1088/0034-4885/75/12/126001} {\bibfield  {journal}
  {\bibinfo  {journal} {Rep. Prog. Phys.}\ }\textbf {\bibinfo {volume} {75}},\
  \bibinfo {pages} {126001} (\bibinfo {year} {2012})}\BibitemShut {NoStop}%
\bibitem [{\citenamefont {Esposito}\ and\ \citenamefont
  {Schaller}(2012)}]{Esposito2012epl}%
  \BibitemOpen
  \bibfield  {author} {\bibinfo {author} {\bibfnamefont {M.}~\bibnamefont
  {Esposito}}\ and\ \bibinfo {author} {\bibfnamefont {G.}~\bibnamefont
  {Schaller}},\ }\bibfield  {title} {\bibinfo {title} {Stochastic
  thermodynamics for {\textquotedblleft}{Maxwell demon}{\textquotedblright}
  feedbacks},\ }\href {https://doi.org/10.1209/0295-5075/99/30003} {\bibfield
  {journal} {\bibinfo  {journal} {Euro phys let}\ }\textbf {\bibinfo {volume}
  {99}},\ \bibinfo {pages} {30003} (\bibinfo {year} {2012})}\BibitemShut
  {NoStop}%
\bibitem [{\citenamefont {Harris}\ and\ \citenamefont
  {Sch{\"u}tz}(2007)}]{Harris2007jsm}%
  \BibitemOpen
  \bibfield  {author} {\bibinfo {author} {\bibfnamefont {R.~J.}\ \bibnamefont
  {Harris}}\ and\ \bibinfo {author} {\bibfnamefont {G.~M.}\ \bibnamefont
  {Sch{\"u}tz}},\ }\bibfield  {title} {\bibinfo {title} {Fluctuation theorems
  for stochastic dynamics},\ }\href
  {https://doi.org/10.1088/1742-5468/2007/07/P07020} {\bibfield  {journal}
  {\bibinfo  {journal} {J. Stat. Mech.}\ }\textbf {\bibinfo {volume} {2007}},\
  \bibinfo {pages} {P07020} (\bibinfo {year} {2007})}\BibitemShut {NoStop}%
\bibitem [{\citenamefont {Jarzynski}(2011)}]{Jarzynski2011arcmp}%
  \BibitemOpen
  \bibfield  {author} {\bibinfo {author} {\bibfnamefont {C.}~\bibnamefont
  {Jarzynski}},\ }\bibfield  {title} {\bibinfo {title} {Equalities and
  inequalities: Irreversibility and the second law of thermodynamics at the
  nanoscale},\ }\href
  {https://doi.org/https://doi.org/10.1146/annurev-conmatphys-062910-140506}
  {\bibfield  {journal} {\bibinfo  {journal} {Annu. Rev. Condens. Matter
  Phys.}\ }\textbf {\bibinfo {volume} {2}},\ \bibinfo {pages} {329} (\bibinfo
  {year} {2011})}\BibitemShut {NoStop}%
\bibitem [{\citenamefont {Horowitz}\ and\ \citenamefont
  {Vaikuntanathan}(2010)}]{Horowitz2010pre}%
  \BibitemOpen
  \bibfield  {author} {\bibinfo {author} {\bibfnamefont {J.~M.}\ \bibnamefont
  {Horowitz}}\ and\ \bibinfo {author} {\bibfnamefont {S.}~\bibnamefont
  {Vaikuntanathan}},\ }\bibfield  {title} {\bibinfo {title} {Nonequilibrium
  detailed fluctuation theorem for repeated discrete feedback},\ }\href
  {https://doi.org/10.1103/PhysRevE.82.061120} {\bibfield  {journal} {\bibinfo
  {journal} {Phys. Rev. E}\ }\textbf {\bibinfo {volume} {82}},\ \bibinfo
  {pages} {061120} (\bibinfo {year} {2010})}\BibitemShut {NoStop}%
\bibitem [{\citenamefont {Abreu}\ and\ \citenamefont
  {Seifert}(2011)}]{Abreu2011epl}%
  \BibitemOpen
  \bibfield  {author} {\bibinfo {author} {\bibfnamefont {D.}~\bibnamefont
  {Abreu}}\ and\ \bibinfo {author} {\bibfnamefont {U.}~\bibnamefont
  {Seifert}},\ }\bibfield  {title} {\bibinfo {title} {Extracting work from a
  single heat bath through feedback},\ }\href
  {https://doi.org/10.1209/0295-5075/94/10001} {\bibfield  {journal} {\bibinfo
  {journal} {Europhys. Lett.}\ }\textbf {\bibinfo {volume} {94}},\ \bibinfo
  {pages} {10001} (\bibinfo {year} {2011})}\BibitemShut {NoStop}%
\bibitem [{\citenamefont {Jun}\ \emph {et~al.}(2014)\citenamefont {Jun},
  \citenamefont {Gavrilov},\ and\ \citenamefont {Bechhoefer}}]{Jun2014prl}%
  \BibitemOpen
  \bibfield  {author} {\bibinfo {author} {\bibfnamefont {Y.}~\bibnamefont
  {Jun}}, \bibinfo {author} {\bibfnamefont {M.}~\bibnamefont {Gavrilov}},\ and\
  \bibinfo {author} {\bibfnamefont {J.}~\bibnamefont {Bechhoefer}},\ }\bibfield
   {title} {\bibinfo {title} {High-{P}recision {T}est of {L}andauer's
  {P}rinciple in a {F}eedback trap},\ }\href
  {https://doi.org/10.1103/PhysRevLett.113.190601} {\bibfield  {journal}
  {\bibinfo  {journal} {Phys. Rev. Lett.}\ }\textbf {\bibinfo {volume} {113}},\
  \bibinfo {pages} {190601} (\bibinfo {year} {2014})}\BibitemShut {NoStop}%
\bibitem [{\citenamefont {Pal}\ \emph {et~al.}(2014)\citenamefont {Pal},
  \citenamefont {Rana}, \citenamefont {Saha},\ and\ \citenamefont
  {Jayannavar}}]{Pal2014pre}%
  \BibitemOpen
  \bibfield  {author} {\bibinfo {author} {\bibfnamefont {P.~S.}\ \bibnamefont
  {Pal}}, \bibinfo {author} {\bibfnamefont {S.}~\bibnamefont {Rana}}, \bibinfo
  {author} {\bibfnamefont {A.}~\bibnamefont {Saha}},\ and\ \bibinfo {author}
  {\bibfnamefont {A.~M.}\ \bibnamefont {Jayannavar}},\ }\bibfield  {title}
  {\bibinfo {title} {Extracting work from a single heat bath: A case study of a
  {B}rownian particle under an external magnetic field in the presence of
  information},\ }\href {https://doi.org/10.1103/PhysRevE.90.022143} {\bibfield
   {journal} {\bibinfo  {journal} {Phys. Rev. E}\ }\textbf {\bibinfo {volume}
  {90}},\ \bibinfo {pages} {022143} (\bibinfo {year} {2014})}\BibitemShut
  {NoStop}%
\bibitem [{\citenamefont {Paneru}\ \emph {et~al.}(2022)\citenamefont {Paneru},
  \citenamefont {Dutta},\ and\ \citenamefont {Pak}}]{paneru2022jpcl}%
  \BibitemOpen
  \bibfield  {author} {\bibinfo {author} {\bibfnamefont {G.}~\bibnamefont
  {Paneru}}, \bibinfo {author} {\bibfnamefont {S.}~\bibnamefont {Dutta}},\ and\
  \bibinfo {author} {\bibfnamefont {H.~K.}\ \bibnamefont {Pak}},\ }\bibfield
  {title} {\bibinfo {title} {Colossal power extraction from active cyclic
  brownian information engines},\ }\href
  {https://doi.org/https://doi.org/10.1021/acs.jpclett.2c01736} {\bibfield
  {journal} {\bibinfo  {journal} {J. Phys. Chem. Lett.}\ }\textbf {\bibinfo
  {volume} {13}},\ \bibinfo {pages} {6912} (\bibinfo {year}
  {2022})}\BibitemShut {NoStop}%
\bibitem [{\citenamefont {Rafeek}\ and\ \citenamefont
  {Mondal}(2024)}]{Rafna2024jcp}%
  \BibitemOpen
  \bibfield  {author} {\bibinfo {author} {\bibfnamefont {R.}~\bibnamefont
  {Rafeek}}\ and\ \bibinfo {author} {\bibfnamefont {D.}~\bibnamefont
  {Mondal}},\ }\bibfield  {title} {\bibinfo {title} {Active {B}rownian
  information engine: Self-propulsion induced colossal performance},\ }\href
  {https://doi.org/https://doi.org/10.1063/5.0229087} {\bibfield  {journal}
  {\bibinfo  {journal} {J. Chem. Phys}\ }\textbf {\bibinfo {volume} {161}},\
  \bibinfo {pages} {124116} (\bibinfo {year} {2024})}\BibitemShut {NoStop}%
\bibitem [{\citenamefont {Rafeek}\ and\ \citenamefont
  {Mondal}(2025{\natexlab{a}})}]{Rafna2025arxiv1}%
  \BibitemOpen
  \bibfield  {author} {\bibinfo {author} {\bibfnamefont {R.}~\bibnamefont
  {Rafeek}}\ and\ \bibinfo {author} {\bibfnamefont {D.}~\bibnamefont
  {Mondal}},\ }\bibfield  {title} {\bibinfo {title} {Performance of a
  {B}rownian information engine through potential profiling: Optimum output
  requisites, heating-to-refrigeration transition and their re-entrance},\
  }\href@noop {} {\bibfield  {journal} {\bibinfo  {journal} {arXiv preprint}\ }
  (\bibinfo {year} {2025}{\natexlab{a}})},\ \Eprint
  {https://arxiv.org/abs/2504.10311} {2504.10311 [cond-mat.soft]} \BibitemShut
  {NoStop}%
\bibitem [{\citenamefont {Saha}\ \emph {et~al.}(2021)\citenamefont {Saha},
  \citenamefont {Lucero}, \citenamefont {Ehrich}, \citenamefont {Sivak},\ and\
  \citenamefont {Bechhoefer}}]{saha2021pnas}%
  \BibitemOpen
  \bibfield  {author} {\bibinfo {author} {\bibfnamefont {T.~K.}\ \bibnamefont
  {Saha}}, \bibinfo {author} {\bibfnamefont {J.~N.~E.}\ \bibnamefont {Lucero}},
  \bibinfo {author} {\bibfnamefont {J.}~\bibnamefont {Ehrich}}, \bibinfo
  {author} {\bibfnamefont {D.~A.}\ \bibnamefont {Sivak}},\ and\ \bibinfo
  {author} {\bibfnamefont {J.}~\bibnamefont {Bechhoefer}},\ }\bibfield  {title}
  {\bibinfo {title} {Maximizing power and velocity of an information engine},\
  }\href {https://www.pnas.org/content/118/20/e2023356118} {\bibfield
  {journal} {\bibinfo  {journal} {Proc. Natl. Acad. Sci. USA}\ }\textbf
  {\bibinfo {volume} {118}} (\bibinfo {year} {2021})}\BibitemShut {NoStop}%
\bibitem [{\citenamefont {Rafeek}\ and\ \citenamefont
  {Mondal}(2025{\natexlab{b}})}]{Rafna2025arxiv2}%
  \BibitemOpen
  \bibfield  {author} {\bibinfo {author} {\bibfnamefont {R.}~\bibnamefont
  {Rafeek}}\ and\ \bibinfo {author} {\bibfnamefont {D.}~\bibnamefont
  {Mondal}},\ }\bibfield  {title} {\bibinfo {title} {Global optimisation of the
  control strategy of a {B}rownian information engine: Efficient
  information-energy exchange in a generalised potential energy surface},\
  }\href@noop {} {\bibfield  {journal} {\bibinfo  {journal} {arXiv preprint}\ }
  (\bibinfo {year} {2025}{\natexlab{b}})},\ \Eprint
  {https://arxiv.org/abs/2505.17546} {2505.17546 [cond-mat.soft]} \BibitemShut
  {NoStop}%
\bibitem [{\citenamefont {Rafeek}\ and\ \citenamefont
  {Mondal}(2025{\natexlab{c}})}]{Rafna2025jpcb}%
  \BibitemOpen
  \bibfield  {author} {\bibinfo {author} {\bibfnamefont {R.}~\bibnamefont
  {Rafeek}}\ and\ \bibinfo {author} {\bibfnamefont {D.}~\bibnamefont
  {Mondal}},\ }\bibfield  {title} {\bibinfo {title} {Achievable
  information-energy exchange in a {B}rownian information engine through
  potential profiling},\ }\href
  {https://doi.org/https://doi.org/10.1021/acs.jpcb.5c00036} {\bibfield
  {journal} {\bibinfo  {journal} {J. Phys. Chem. B}\ }\textbf {\bibinfo
  {volume} {129}},\ \bibinfo {pages} {2971} (\bibinfo {year}
  {2025}{\natexlab{c}})}\BibitemShut {NoStop}%
\bibitem [{\citenamefont {Kim}\ \emph {et~al.}(2011)\citenamefont {Kim},
  \citenamefont {Sagawa}, \citenamefont {De~Liberato},\ and\ \citenamefont
  {Ueda}}]{Kim2011prl}%
  \BibitemOpen
  \bibfield  {author} {\bibinfo {author} {\bibfnamefont {S.~W.}\ \bibnamefont
  {Kim}}, \bibinfo {author} {\bibfnamefont {T.}~\bibnamefont {Sagawa}},
  \bibinfo {author} {\bibfnamefont {S.}~\bibnamefont {De~Liberato}},\ and\
  \bibinfo {author} {\bibfnamefont {M.}~\bibnamefont {Ueda}},\ }\bibfield
  {title} {\bibinfo {title} {Quantum {S}zilard {E}ngine},\ }\href
  {https://doi.org/10.1103/PhysRevLett.106.070401} {\bibfield  {journal}
  {\bibinfo  {journal} {Phys. Rev. Lett.}\ }\textbf {\bibinfo {volume} {106}},\
  \bibinfo {pages} {070401} (\bibinfo {year} {2011})}\BibitemShut {NoStop}%
\bibitem [{\citenamefont {Bruschi}\ \emph {et~al.}(2015)\citenamefont
  {Bruschi}, \citenamefont {Perarnau-Llobet}, \citenamefont {Friis},
  \citenamefont {Hovhannisyan},\ and\ \citenamefont {Huber}}]{Bruschi2015pre}%
  \BibitemOpen
  \bibfield  {author} {\bibinfo {author} {\bibfnamefont {D.~E.}\ \bibnamefont
  {Bruschi}}, \bibinfo {author} {\bibfnamefont {M.}~\bibnamefont
  {Perarnau-Llobet}}, \bibinfo {author} {\bibfnamefont {N.}~\bibnamefont
  {Friis}}, \bibinfo {author} {\bibfnamefont {K.~V.}\ \bibnamefont
  {Hovhannisyan}},\ and\ \bibinfo {author} {\bibfnamefont {M.}~\bibnamefont
  {Huber}},\ }\bibfield  {title} {\bibinfo {title} {Thermodynamics of creating
  correlations: Limitations and optimal protocols},\ }\href
  {https://doi.org/10.1103/PhysRevE.91.032118} {\bibfield  {journal} {\bibinfo
  {journal} {Phys. Rev. E}\ }\textbf {\bibinfo {volume} {91}},\ \bibinfo
  {pages} {032118} (\bibinfo {year} {2015})}\BibitemShut {NoStop}%
\bibitem [{\citenamefont {Goold}\ \emph {et~al.}(2016)\citenamefont {Goold},
  \citenamefont {Huber}, \citenamefont {Riera}, \citenamefont {Del~Rio},\ and\
  \citenamefont {Skrzypczyk}}]{Goold2016jphysA}%
  \BibitemOpen
  \bibfield  {author} {\bibinfo {author} {\bibfnamefont {J.}~\bibnamefont
  {Goold}}, \bibinfo {author} {\bibfnamefont {M.}~\bibnamefont {Huber}},
  \bibinfo {author} {\bibfnamefont {A.}~\bibnamefont {Riera}}, \bibinfo
  {author} {\bibfnamefont {L.}~\bibnamefont {Del~Rio}},\ and\ \bibinfo {author}
  {\bibfnamefont {P.}~\bibnamefont {Skrzypczyk}},\ }\bibfield  {title}
  {\bibinfo {title} {The role of quantum information in thermodynamics—a
  topical review},\ }\href {https://doi.org/10.1088/1751-8113/49/14/143001}
  {\bibfield  {journal} {\bibinfo  {journal} {J. Phys. A}\ }\textbf {\bibinfo
  {volume} {49}},\ \bibinfo {pages} {143001} (\bibinfo {year}
  {2016})}\BibitemShut {NoStop}%
\bibitem [{\citenamefont {Masuyama}\ \emph {et~al.}(2018)\citenamefont
  {Masuyama}, \citenamefont {Funo}, \citenamefont {Murashita}, \citenamefont
  {Noguchi}, \citenamefont {Kono}, \citenamefont {Tabuchi}, \citenamefont
  {Yamazaki}, \citenamefont {Ueda},\ and\ \citenamefont
  {Nakamura}}]{Masuyama2018natcom}%
  \BibitemOpen
  \bibfield  {author} {\bibinfo {author} {\bibfnamefont {Y.}~\bibnamefont
  {Masuyama}}, \bibinfo {author} {\bibfnamefont {K.}~\bibnamefont {Funo}},
  \bibinfo {author} {\bibfnamefont {Y.}~\bibnamefont {Murashita}}, \bibinfo
  {author} {\bibfnamefont {A.}~\bibnamefont {Noguchi}}, \bibinfo {author}
  {\bibfnamefont {S.}~\bibnamefont {Kono}}, \bibinfo {author} {\bibfnamefont
  {Y.}~\bibnamefont {Tabuchi}}, \bibinfo {author} {\bibfnamefont
  {R.}~\bibnamefont {Yamazaki}}, \bibinfo {author} {\bibfnamefont
  {M.}~\bibnamefont {Ueda}},\ and\ \bibinfo {author} {\bibfnamefont
  {Y.}~\bibnamefont {Nakamura}},\ }\bibfield  {title} {\bibinfo {title}
  {Information-to-work conversion by {M}axwell’s demon in a superconducting
  circuit quantum electrodynamical system},\ }\href
  {https://doi.org/https://doi.org/10.1038/s41467-018-03686-y} {\bibfield
  {journal} {\bibinfo  {journal} {Nat Commun}\ }\textbf {\bibinfo {volume}
  {9}},\ \bibinfo {pages} {1291} (\bibinfo {year} {2018})}\BibitemShut
  {NoStop}%
\bibitem [{\citenamefont {Naghiloo}\ \emph {et~al.}(2018)\citenamefont
  {Naghiloo}, \citenamefont {Alonso}, \citenamefont {Romito}, \citenamefont
  {Lutz},\ and\ \citenamefont {Murch}}]{Naghilo2018prl}%
  \BibitemOpen
  \bibfield  {author} {\bibinfo {author} {\bibfnamefont {M.}~\bibnamefont
  {Naghiloo}}, \bibinfo {author} {\bibfnamefont {J.~J.}\ \bibnamefont
  {Alonso}}, \bibinfo {author} {\bibfnamefont {A.}~\bibnamefont {Romito}},
  \bibinfo {author} {\bibfnamefont {E.}~\bibnamefont {Lutz}},\ and\ \bibinfo
  {author} {\bibfnamefont {K.~W.}\ \bibnamefont {Murch}},\ }\bibfield  {title}
  {\bibinfo {title} {Information gain and loss for a quantum maxwell's demon},\
  }\href {https://doi.org/10.1103/PhysRevLett.121.030604} {\bibfield  {journal}
  {\bibinfo  {journal} {Phys. Rev. Lett.}\ }\textbf {\bibinfo {volume} {121}},\
  \bibinfo {pages} {030604} (\bibinfo {year} {2018})}\BibitemShut {NoStop}%
\bibitem [{\citenamefont {Koski}\ \emph
  {et~al.}(2014{\natexlab{a}})\citenamefont {Koski}, \citenamefont {Maisi},
  \citenamefont {Pekola},\ and\ \citenamefont {Averin}}]{Koski2014pnas}%
  \BibitemOpen
  \bibfield  {author} {\bibinfo {author} {\bibfnamefont {J.~V.}\ \bibnamefont
  {Koski}}, \bibinfo {author} {\bibfnamefont {V.~F.}\ \bibnamefont {Maisi}},
  \bibinfo {author} {\bibfnamefont {J.~P.}\ \bibnamefont {Pekola}},\ and\
  \bibinfo {author} {\bibfnamefont {D.~V.}\ \bibnamefont {Averin}},\ }\bibfield
   {title} {\bibinfo {title} {Experimental realization of a {S}zilard engine
  with a single electron},\ }\href {https://doi.org/10.1073/pnas.1406966111}
  {\bibfield  {journal} {\bibinfo  {journal} {Proc. Natl. Acad. Sci. U. S. A}\
  }\textbf {\bibinfo {volume} {111}},\ \bibinfo {pages} {13786} (\bibinfo
  {year} {2014}{\natexlab{a}})}\BibitemShut {NoStop}%
\bibitem [{\citenamefont {Koski}\ \emph
  {et~al.}(2014{\natexlab{b}})\citenamefont {Koski}, \citenamefont {Maisi},
  \citenamefont {Sagawa},\ and\ \citenamefont {Pekola}}]{Koski2014prl}%
  \BibitemOpen
  \bibfield  {author} {\bibinfo {author} {\bibfnamefont {J.~V.}\ \bibnamefont
  {Koski}}, \bibinfo {author} {\bibfnamefont {V.~F.}\ \bibnamefont {Maisi}},
  \bibinfo {author} {\bibfnamefont {T.}~\bibnamefont {Sagawa}},\ and\ \bibinfo
  {author} {\bibfnamefont {J.~P.}\ \bibnamefont {Pekola}},\ }\bibfield  {title}
  {\bibinfo {title} {Experimental {O}bservation of the {R}ole of {M}utual
  {I}nformation in the {N}onequilibrium {D}ynamics of a {M}axwell {D}emon},\
  }\href {https://doi.org/10.1103/PhysRevLett.113.030601} {\bibfield  {journal}
  {\bibinfo  {journal} {Phys. Rev. Lett.}\ }\textbf {\bibinfo {volume} {113}},\
  \bibinfo {pages} {030601} (\bibinfo {year} {2014}{\natexlab{b}})}\BibitemShut
  {NoStop}%
\bibitem [{\citenamefont {Koski}\ \emph {et~al.}(2015)\citenamefont {Koski},
  \citenamefont {Kutvonen}, \citenamefont {Khaymovich}, \citenamefont
  {Ala-Nissila},\ and\ \citenamefont {Pekola}}]{Koski2015prl}%
  \BibitemOpen
  \bibfield  {author} {\bibinfo {author} {\bibfnamefont {J.~V.}\ \bibnamefont
  {Koski}}, \bibinfo {author} {\bibfnamefont {A.}~\bibnamefont {Kutvonen}},
  \bibinfo {author} {\bibfnamefont {I.~M.}\ \bibnamefont {Khaymovich}},
  \bibinfo {author} {\bibfnamefont {T.}~\bibnamefont {Ala-Nissila}},\ and\
  \bibinfo {author} {\bibfnamefont {J.~P.}\ \bibnamefont {Pekola}},\ }\bibfield
   {title} {\bibinfo {title} {On-chip maxwell's demon as an information-powered
  refrigerator},\ }\href {https://doi.org/10.1103/PhysRevLett.115.260602}
  {\bibfield  {journal} {\bibinfo  {journal} {Phys. Rev. Lett.}\ }\textbf
  {\bibinfo {volume} {115}},\ \bibinfo {pages} {260602} (\bibinfo {year}
  {2015})}\BibitemShut {NoStop}%
\bibitem [{\citenamefont {Kutvonen}\ \emph {et~al.}(2016)\citenamefont
  {Kutvonen}, \citenamefont {Koski},\ and\ \citenamefont
  {Ala-Nissila}}]{Kutvonen2016Scirep}%
  \BibitemOpen
  \bibfield  {author} {\bibinfo {author} {\bibfnamefont {A.}~\bibnamefont
  {Kutvonen}}, \bibinfo {author} {\bibfnamefont {J.}~\bibnamefont {Koski}},\
  and\ \bibinfo {author} {\bibfnamefont {T.}~\bibnamefont {Ala-Nissila}},\
  }\bibfield  {title} {\bibinfo {title} {Thermodynamics and efficiency of an
  autonomous on-chip maxwell's demon},\ }\href
  {https://doi.org/10.1038/srep21126} {\bibfield  {journal} {\bibinfo
  {journal} {Sci Rep}\ }\textbf {\bibinfo {volume} {6}},\ \bibinfo {pages}
  {21126} (\bibinfo {year} {2016})}\BibitemShut {NoStop}%
\bibitem [{\citenamefont {Lopez}\ \emph {et~al.}(2008)\citenamefont {Lopez},
  \citenamefont {Kuwada}, \citenamefont {Craig}, \citenamefont {Long},\ and\
  \citenamefont {Linke}}]{Lopez2008prl}%
  \BibitemOpen
  \bibfield  {author} {\bibinfo {author} {\bibfnamefont {B.~J.}\ \bibnamefont
  {Lopez}}, \bibinfo {author} {\bibfnamefont {N.~J.}\ \bibnamefont {Kuwada}},
  \bibinfo {author} {\bibfnamefont {E.~M.}\ \bibnamefont {Craig}}, \bibinfo
  {author} {\bibfnamefont {B.~R.}\ \bibnamefont {Long}},\ and\ \bibinfo
  {author} {\bibfnamefont {H.}~\bibnamefont {Linke}},\ }\bibfield  {title}
  {\bibinfo {title} {Realization of a {F}eedback {C}ontrolled {F}lashing
  {R}atchet},\ }\href {https://doi.org/10.1103/PhysRevLett.101.220601}
  {\bibfield  {journal} {\bibinfo  {journal} {Phys. Rev. Lett.}\ }\textbf
  {\bibinfo {volume} {101}},\ \bibinfo {pages} {220601} (\bibinfo {year}
  {2008})}\BibitemShut {NoStop}%
\bibitem [{\citenamefont {Toyabe}\ \emph {et~al.}(2010)\citenamefont {Toyabe},
  \citenamefont {Sagawa}, \citenamefont {Ueda}, \citenamefont {Muneyuki},\ and\
  \citenamefont {Sano}}]{Toyabe2010natphys}%
  \BibitemOpen
  \bibfield  {author} {\bibinfo {author} {\bibfnamefont {S.}~\bibnamefont
  {Toyabe}}, \bibinfo {author} {\bibfnamefont {T.}~\bibnamefont {Sagawa}},
  \bibinfo {author} {\bibfnamefont {M.}~\bibnamefont {Ueda}}, \bibinfo {author}
  {\bibfnamefont {E.}~\bibnamefont {Muneyuki}},\ and\ \bibinfo {author}
  {\bibfnamefont {M.}~\bibnamefont {Sano}},\ }\bibfield  {title} {\bibinfo
  {title} {Experimental demonstration of information-to-energy conversion and
  validation of the generalized {J}arzynski equality},\ }\href
  {https://doi.org/10.1038/nphys1821} {\bibfield  {journal} {\bibinfo
  {journal} {Nat. Phys.}\ }\textbf {\bibinfo {volume} {6}},\ \bibinfo {pages}
  {988} (\bibinfo {year} {2010})}\BibitemShut {NoStop}%
\bibitem [{\citenamefont {Esposito}\ and\ \citenamefont {Van~den
  Broeck}(2011)}]{Esposito2011epl}%
  \BibitemOpen
  \bibfield  {author} {\bibinfo {author} {\bibfnamefont {M.}~\bibnamefont
  {Esposito}}\ and\ \bibinfo {author} {\bibfnamefont {C.}~\bibnamefont {Van~den
  Broeck}},\ }\bibfield  {title} {\bibinfo {title} {Second law and {L}andauer
  principle far from equilibrium},\ }\href
  {https://doi.org/10.1209/0295-5075/95/40004} {\bibfield  {journal} {\bibinfo
  {journal} {Europhys. Lett.}\ }\textbf {\bibinfo {volume} {95}},\ \bibinfo
  {pages} {40004} (\bibinfo {year} {2011})}\BibitemShut {NoStop}%
\bibitem [{\citenamefont {Still}\ \emph {et~al.}(2012)\citenamefont {Still},
  \citenamefont {Sivak}, \citenamefont {Bell},\ and\ \citenamefont
  {Crooks}}]{Still2012prl}%
  \BibitemOpen
  \bibfield  {author} {\bibinfo {author} {\bibfnamefont {S.}~\bibnamefont
  {Still}}, \bibinfo {author} {\bibfnamefont {D.~A.}\ \bibnamefont {Sivak}},
  \bibinfo {author} {\bibfnamefont {A.~J.}\ \bibnamefont {Bell}},\ and\
  \bibinfo {author} {\bibfnamefont {G.~E.}\ \bibnamefont {Crooks}},\ }\bibfield
   {title} {\bibinfo {title} {Thermodynamics of {P}rediction},\ }\href
  {https://doi.org/10.1103/PhysRevLett.109.120604} {\bibfield  {journal}
  {\bibinfo  {journal} {Phys. Rev. Lett.}\ }\textbf {\bibinfo {volume} {109}},\
  \bibinfo {pages} {120604} (\bibinfo {year} {2012})}\BibitemShut {NoStop}%
\bibitem [{\citenamefont {Abreu}\ and\ \citenamefont
  {Seifert}(2012)}]{Abreu2012prl}%
  \BibitemOpen
  \bibfield  {author} {\bibinfo {author} {\bibfnamefont {D.}~\bibnamefont
  {Abreu}}\ and\ \bibinfo {author} {\bibfnamefont {U.}~\bibnamefont
  {Seifert}},\ }\bibfield  {title} {\bibinfo {title} {Thermodynamics of
  {G}enuine {N}onequilibrium {S}tates under {F}eedback {C}ontrol},\ }\href
  {https://doi.org/10.1103/PhysRevLett.108.030601} {\bibfield  {journal}
  {\bibinfo  {journal} {Phys. Rev. Lett.}\ }\textbf {\bibinfo {volume} {108}},\
  \bibinfo {pages} {030601} (\bibinfo {year} {2012})}\BibitemShut {NoStop}%
\bibitem [{\citenamefont {Bauer}\ \emph {et~al.}(2012)\citenamefont {Bauer},
  \citenamefont {Abreu},\ and\ \citenamefont {Seifert}}]{Bauer2012jphysa}%
  \BibitemOpen
  \bibfield  {author} {\bibinfo {author} {\bibfnamefont {M.}~\bibnamefont
  {Bauer}}, \bibinfo {author} {\bibfnamefont {D.}~\bibnamefont {Abreu}},\ and\
  \bibinfo {author} {\bibfnamefont {U.}~\bibnamefont {Seifert}},\ }\bibfield
  {title} {\bibinfo {title} {Efficiency of a {B}rownian information machine},\
  }\href {https://doi.org/10.1088/1751-8113/45/16/162001} {\bibfield  {journal}
  {\bibinfo  {journal} {J. Phys. A: Math. Theor}\ }\textbf {\bibinfo {volume}
  {45}},\ \bibinfo {pages} {162001} (\bibinfo {year} {2012})}\BibitemShut
  {NoStop}%
\bibitem [{\citenamefont {Park}\ \emph {et~al.}(2016)\citenamefont {Park},
  \citenamefont {Lee},\ and\ \citenamefont {Noh}}]{Park2016pre}%
  \BibitemOpen
  \bibfield  {author} {\bibinfo {author} {\bibfnamefont {J.~M.}\ \bibnamefont
  {Park}}, \bibinfo {author} {\bibfnamefont {J.~S.}\ \bibnamefont {Lee}},\ and\
  \bibinfo {author} {\bibfnamefont {J.~D.}\ \bibnamefont {Noh}},\ }\bibfield
  {title} {\bibinfo {title} {Optimal tuning of a confined {B}rownian
  information engine},\ }\href {https://doi.org/10.1103/PhysRevE.93.032146}
  {\bibfield  {journal} {\bibinfo  {journal} {Phys. Rev. E}\ }\textbf {\bibinfo
  {volume} {93}},\ \bibinfo {pages} {032146} (\bibinfo {year}
  {2016})}\BibitemShut {NoStop}%
\bibitem [{\citenamefont {Paneru}\ \emph
  {et~al.}(2018{\natexlab{b}})\citenamefont {Paneru}, \citenamefont {Lee},
  \citenamefont {Park}, \citenamefont {Park}, \citenamefont {Noh},\ and\
  \citenamefont {Pak}}]{Paneru2018pre}%
  \BibitemOpen
  \bibfield  {author} {\bibinfo {author} {\bibfnamefont {G.}~\bibnamefont
  {Paneru}}, \bibinfo {author} {\bibfnamefont {D.~Y.}\ \bibnamefont {Lee}},
  \bibinfo {author} {\bibfnamefont {J.-M.}\ \bibnamefont {Park}}, \bibinfo
  {author} {\bibfnamefont {J.~T.}\ \bibnamefont {Park}}, \bibinfo {author}
  {\bibfnamefont {J.~D.}\ \bibnamefont {Noh}},\ and\ \bibinfo {author}
  {\bibfnamefont {H.~K.}\ \bibnamefont {Pak}},\ }\bibfield  {title} {\bibinfo
  {title} {Optimal tuning of a {B}rownian information engine operating in a
  nonequilibrium steady state},\ }\href
  {https://doi.org/10.1103/PhysRevE.98.052119} {\bibfield  {journal} {\bibinfo
  {journal} {Phys. Rev. E}\ }\textbf {\bibinfo {volume} {98}},\ \bibinfo
  {pages} {052119} (\bibinfo {year} {2018}{\natexlab{b}})}\BibitemShut
  {NoStop}%
\bibitem [{\citenamefont {Lee}\ \emph {et~al.}(2018)\citenamefont {Lee},
  \citenamefont {Um}, \citenamefont {Paneru},\ and\ \citenamefont
  {Pak}}]{Lee2018scirep}%
  \BibitemOpen
  \bibfield  {author} {\bibinfo {author} {\bibfnamefont {D.~Y.}\ \bibnamefont
  {Lee}}, \bibinfo {author} {\bibfnamefont {J.}~\bibnamefont {Um}}, \bibinfo
  {author} {\bibfnamefont {G.}~\bibnamefont {Paneru}},\ and\ \bibinfo {author}
  {\bibfnamefont {H.~K.}\ \bibnamefont {Pak}},\ }\bibfield  {title} {\bibinfo
  {title} {An experimentally-achieved information-driven {B}rownian motor shows
  maximum power at the relaxation time},\ }\href
  {https://doi.org/https://doi.org/10.1038/s41598-018-30495-6} {\bibfield
  {journal} {\bibinfo  {journal} {Sci Rep}\ }\textbf {\bibinfo {volume} {8}},\
  \bibinfo {pages} {12121} (\bibinfo {year} {2018})}\BibitemShut {NoStop}%
\bibitem [{\citenamefont {Paneru}\ \emph {et~al.}(2020)\citenamefont {Paneru},
  \citenamefont {Dutta}, \citenamefont {Sagawa}, \citenamefont {Tlusty},\ and\
  \citenamefont {Pak}}]{Paneru2020Natcommun}%
  \BibitemOpen
  \bibfield  {author} {\bibinfo {author} {\bibfnamefont {G.}~\bibnamefont
  {Paneru}}, \bibinfo {author} {\bibfnamefont {S.}~\bibnamefont {Dutta}},
  \bibinfo {author} {\bibfnamefont {T.}~\bibnamefont {Sagawa}}, \bibinfo
  {author} {\bibfnamefont {T.}~\bibnamefont {Tlusty}},\ and\ \bibinfo {author}
  {\bibfnamefont {H.~K.}\ \bibnamefont {Pak}},\ }\bibfield  {title} {\bibinfo
  {title} {Efficiency fluctuations and noise induced refrigerator-to-heater
  transition in information engines},\ }\href
  {https://doi.org/10.1038/s41467-020-14823-x} {\bibfield  {journal} {\bibinfo
  {journal} {Nat. Commun.}\ }\textbf {\bibinfo {volume} {11}},\ \bibinfo
  {pages} {1012} (\bibinfo {year} {2020})}\BibitemShut {NoStop}%
\bibitem [{\citenamefont {Zwanzig}(1992)}]{Zwanzig1992jpc}%
  \BibitemOpen
  \bibfield  {author} {\bibinfo {author} {\bibfnamefont {R.}~\bibnamefont
  {Zwanzig}},\ }\bibfield  {title} {\bibinfo {title} {Diffusion past an entropy
  barrier},\ }\href {https://doi.org/10.1021/j100189a004} {\bibfield  {journal}
  {\bibinfo  {journal} {J. Phys. Chem.}\ }\textbf {\bibinfo {volume} {96}},\
  \bibinfo {pages} {3926} (\bibinfo {year} {1992})}\BibitemShut {NoStop}%
\bibitem [{\citenamefont {Reguera}\ and\ \citenamefont
  {Rubi}(2001)}]{Reguera2001pre}%
  \BibitemOpen
  \bibfield  {author} {\bibinfo {author} {\bibfnamefont {D.}~\bibnamefont
  {Reguera}}\ and\ \bibinfo {author} {\bibfnamefont {J.}~\bibnamefont {Rubi}},\
  }\bibfield  {title} {\bibinfo {title} {Kinetic equations for diffusion in the
  presence of entropic barriers},\ }\href
  {https://doi.org/10.1103/PhysRevE.64.061106} {\bibfield  {journal} {\bibinfo
  {journal} {Phys. Rev. E}\ }\textbf {\bibinfo {volume} {64}},\ \bibinfo
  {pages} {061106} (\bibinfo {year} {2001})}\BibitemShut {NoStop}%
\bibitem [{\citenamefont {Reguera}\ \emph {et~al.}(2006)\citenamefont
  {Reguera}, \citenamefont {Schmid}, \citenamefont {Burada}, \citenamefont
  {Rubi}, \citenamefont {Reimann},\ and\ \citenamefont
  {H{\"a}nggi}}]{Reguera2006prl}%
  \BibitemOpen
  \bibfield  {author} {\bibinfo {author} {\bibfnamefont {D.}~\bibnamefont
  {Reguera}}, \bibinfo {author} {\bibfnamefont {G.}~\bibnamefont {Schmid}},
  \bibinfo {author} {\bibfnamefont {P.~S.}\ \bibnamefont {Burada}}, \bibinfo
  {author} {\bibfnamefont {J.}~\bibnamefont {Rubi}}, \bibinfo {author}
  {\bibfnamefont {P.}~\bibnamefont {Reimann}},\ and\ \bibinfo {author}
  {\bibfnamefont {P.}~\bibnamefont {H{\"a}nggi}},\ }\bibfield  {title}
  {\bibinfo {title} {Entropic {T}ransport: {K}inetics, {S}caling, and {C}ontrol
  {M}echanisms},\ }\href {https://doi.org/10.1103/PhysRevLett.96.130603}
  {\bibfield  {journal} {\bibinfo  {journal} {Phys. Rev. Lett.}\ }\textbf
  {\bibinfo {volume} {96}},\ \bibinfo {pages} {130603} (\bibinfo {year}
  {2006})}\BibitemShut {NoStop}%
\bibitem [{\citenamefont {Burada}\ \emph
  {et~al.}(2009{\natexlab{a}})\citenamefont {Burada}, \citenamefont
  {H{\"a}nggi}, \citenamefont {Marchesoni}, \citenamefont {Schmid},\ and\
  \citenamefont {Talkner}}]{Burada2009cpc}%
  \BibitemOpen
  \bibfield  {author} {\bibinfo {author} {\bibfnamefont {P.~S.}\ \bibnamefont
  {Burada}}, \bibinfo {author} {\bibfnamefont {P.}~\bibnamefont {H{\"a}nggi}},
  \bibinfo {author} {\bibfnamefont {F.}~\bibnamefont {Marchesoni}}, \bibinfo
  {author} {\bibfnamefont {G.}~\bibnamefont {Schmid}},\ and\ \bibinfo {author}
  {\bibfnamefont {P.}~\bibnamefont {Talkner}},\ }\bibfield  {title} {\bibinfo
  {title} {Diffusion in confined geometries},\ }\href
  {https://doi.org/https://doi.org/10.1002/cphc.200800526} {\bibfield
  {journal} {\bibinfo  {journal} {ChemPhysChem}\ }\textbf {\bibinfo {volume}
  {10}},\ \bibinfo {pages} {45} (\bibinfo {year}
  {2009}{\natexlab{a}})}\BibitemShut {NoStop}%
\bibitem [{\citenamefont {Burada}\ \emph {et~al.}(2007)\citenamefont {Burada},
  \citenamefont {Schmid}, \citenamefont {Reguera}, \citenamefont {Rubi},\ and\
  \citenamefont {H{\"a}nggi}}]{Burada2007pre}%
  \BibitemOpen
  \bibfield  {author} {\bibinfo {author} {\bibfnamefont {P.~S.}\ \bibnamefont
  {Burada}}, \bibinfo {author} {\bibfnamefont {G.}~\bibnamefont {Schmid}},
  \bibinfo {author} {\bibfnamefont {D.}~\bibnamefont {Reguera}}, \bibinfo
  {author} {\bibfnamefont {J.}~\bibnamefont {Rubi}},\ and\ \bibinfo {author}
  {\bibfnamefont {P.}~\bibnamefont {H{\"a}nggi}},\ }\bibfield  {title}
  {\bibinfo {title} {Biased diffusion in confined media: {T}est of the
  {F}ick-{J}acobs approximation and validity criteria},\ }\href
  {https://doi.org/10.1103/PhysRevE.75.051111} {\bibfield  {journal} {\bibinfo
  {journal} {Phys. Rev. E}\ }\textbf {\bibinfo {volume} {75}},\ \bibinfo
  {pages} {051111} (\bibinfo {year} {2007})}\BibitemShut {NoStop}%
\bibitem [{\citenamefont {Burada}\ \emph
  {et~al.}(2008{\natexlab{a}})\citenamefont {Burada}, \citenamefont {Schmid},
  \citenamefont {Reguera}, \citenamefont {Vainstein}, \citenamefont {Rubi},\
  and\ \citenamefont {H{\"a}nggi}}]{Burada2008prl}%
  \BibitemOpen
  \bibfield  {author} {\bibinfo {author} {\bibfnamefont {P.~S.}\ \bibnamefont
  {Burada}}, \bibinfo {author} {\bibfnamefont {G.}~\bibnamefont {Schmid}},
  \bibinfo {author} {\bibfnamefont {D.}~\bibnamefont {Reguera}}, \bibinfo
  {author} {\bibfnamefont {M.~H.}\ \bibnamefont {Vainstein}}, \bibinfo {author}
  {\bibfnamefont {J.}~\bibnamefont {Rubi}},\ and\ \bibinfo {author}
  {\bibfnamefont {P.}~\bibnamefont {H{\"a}nggi}},\ }\bibfield  {title}
  {\bibinfo {title} {Entropic {S}tochastic {R}esonance},\ }\href
  {https://doi.org/10.1103/PhysRevLett.101.130602} {\bibfield  {journal}
  {\bibinfo  {journal} {Phys. Rev. Lett.}\ }\textbf {\bibinfo {volume} {101}},\
  \bibinfo {pages} {130602} (\bibinfo {year} {2008}{\natexlab{a}})}\BibitemShut
  {NoStop}%
\bibitem [{\citenamefont {Mondal}\ and\ \citenamefont
  {Ray}(2010)}]{Mondal2010pre}%
  \BibitemOpen
  \bibfield  {author} {\bibinfo {author} {\bibfnamefont {D.}~\bibnamefont
  {Mondal}}\ and\ \bibinfo {author} {\bibfnamefont {D.~S.}\ \bibnamefont
  {Ray}},\ }\bibfield  {title} {\bibinfo {title} {Diffusion over an entropic
  barrier: {N}on-{A}rrhenius behavior},\ }\href
  {https://doi.org/10.1103/PhysRevE.82.032103} {\bibfield  {journal} {\bibinfo
  {journal} {Phys. Rev. E}\ }\textbf {\bibinfo {volume} {82}},\ \bibinfo
  {pages} {032103} (\bibinfo {year} {2010})}\BibitemShut {NoStop}%
\bibitem [{\citenamefont {Mondal}\ \emph
  {et~al.}(2010{\natexlab{a}})\citenamefont {Mondal}, \citenamefont {Das},\
  and\ \citenamefont {Ray}}]{Mondal2010jcp1}%
  \BibitemOpen
  \bibfield  {author} {\bibinfo {author} {\bibfnamefont {D.}~\bibnamefont
  {Mondal}}, \bibinfo {author} {\bibfnamefont {M.}~\bibnamefont {Das}},\ and\
  \bibinfo {author} {\bibfnamefont {D.~S.}\ \bibnamefont {Ray}},\ }\bibfield
  {title} {\bibinfo {title} {Entropic noise-induced nonequilibrium
  transition},\ }\href {https://doi.org/10.1063/1.3505454} {\bibfield
  {journal} {\bibinfo  {journal} {J. Chem. Phys.}\ }\textbf {\bibinfo {volume}
  {133}},\ \bibinfo {pages} {204102} (\bibinfo {year}
  {2010}{\natexlab{a}})}\BibitemShut {NoStop}%
\bibitem [{\citenamefont {Pagliara}\ \emph
  {et~al.}(2014{\natexlab{a}})\citenamefont {Pagliara}, \citenamefont
  {Dettmer},\ and\ \citenamefont {Keyser}}]{pagliara2014prl}%
  \BibitemOpen
  \bibfield  {author} {\bibinfo {author} {\bibfnamefont {S.}~\bibnamefont
  {Pagliara}}, \bibinfo {author} {\bibfnamefont {S.~L.}\ \bibnamefont
  {Dettmer}},\ and\ \bibinfo {author} {\bibfnamefont {U.~F.}\ \bibnamefont
  {Keyser}},\ }\bibfield  {title} {\bibinfo {title} {Channel-facilitated
  diffusion boosted by particle binding at the channel entrance},\ }\href
  {https://doi.org/10.1103/PhysRevLett.113.048102} {\bibfield  {journal}
  {\bibinfo  {journal} {Phys. Rev. Lett.}\ }\textbf {\bibinfo {volume} {113}},\
  \bibinfo {pages} {048102} (\bibinfo {year} {2014}{\natexlab{a}})}\BibitemShut
  {NoStop}%
\bibitem [{\citenamefont {Mondal}\ and\ \citenamefont
  {Ray}(2011)}]{Mondal2011jcp}%
  \BibitemOpen
  \bibfield  {author} {\bibinfo {author} {\bibfnamefont {D.}~\bibnamefont
  {Mondal}}\ and\ \bibinfo {author} {\bibfnamefont {D.~S.}\ \bibnamefont
  {Ray}},\ }\bibfield  {title} {\bibinfo {title} {Asymmetric stochastic
  localization in geometry controlled kinetics},\ }\href
  {https://doi.org/10.1063/1.3658486} {\bibfield  {journal} {\bibinfo
  {journal} {J. Chem. Phys.}\ }\textbf {\bibinfo {volume} {135}},\ \bibinfo
  {pages} {194111} (\bibinfo {year} {2011})}\BibitemShut {NoStop}%
\bibitem [{\citenamefont {Marchesoni}\ and\ \citenamefont
  {Savel'ev}(2009)}]{Marchesoni2009pre}%
  \BibitemOpen
  \bibfield  {author} {\bibinfo {author} {\bibfnamefont {F.}~\bibnamefont
  {Marchesoni}}\ and\ \bibinfo {author} {\bibfnamefont {S.}~\bibnamefont
  {Savel'ev}},\ }\bibfield  {title} {\bibinfo {title} {Rectification currents
  in two-dimensional artificial channels},\ }\href
  {https://doi.org/10.1103/PhysRevE.80.011120} {\bibfield  {journal} {\bibinfo
  {journal} {Phys. Rev. E}\ }\textbf {\bibinfo {volume} {80}},\ \bibinfo
  {pages} {011120} (\bibinfo {year} {2009})}\BibitemShut {NoStop}%
\bibitem [{\citenamefont {Mondal}\ \emph
  {et~al.}(2010{\natexlab{b}})\citenamefont {Mondal}, \citenamefont {Das},\
  and\ \citenamefont {Ray}}]{Mondal2010jcp2}%
  \BibitemOpen
  \bibfield  {author} {\bibinfo {author} {\bibfnamefont {D.}~\bibnamefont
  {Mondal}}, \bibinfo {author} {\bibfnamefont {M.}~\bibnamefont {Das}},\ and\
  \bibinfo {author} {\bibfnamefont {D.~S.}\ \bibnamefont {Ray}},\ }\bibfield
  {title} {\bibinfo {title} {Entropic resonant activation},\ }\href
  {https://doi.org/10.1063/1.3431042} {\bibfield  {journal} {\bibinfo
  {journal} {J. Chem. Phys.}\ }\textbf {\bibinfo {volume} {132}},\ \bibinfo
  {pages} {224102} (\bibinfo {year} {2010}{\natexlab{b}})}\BibitemShut
  {NoStop}%
\bibitem [{\citenamefont {Burada}\ \emph
  {et~al.}(2008{\natexlab{b}})\citenamefont {Burada}, \citenamefont {Schmid},
  \citenamefont {Talkner}, \citenamefont {H{\"a}nggi}, \citenamefont
  {Reguera},\ and\ \citenamefont {Rub{\'\i}}}]{Burada2008biosyst}%
  \BibitemOpen
  \bibfield  {author} {\bibinfo {author} {\bibfnamefont {P.~S.}\ \bibnamefont
  {Burada}}, \bibinfo {author} {\bibfnamefont {G.}~\bibnamefont {Schmid}},
  \bibinfo {author} {\bibfnamefont {P.}~\bibnamefont {Talkner}}, \bibinfo
  {author} {\bibfnamefont {P.}~\bibnamefont {H{\"a}nggi}}, \bibinfo {author}
  {\bibfnamefont {D.}~\bibnamefont {Reguera}},\ and\ \bibinfo {author}
  {\bibfnamefont {J.~M.}\ \bibnamefont {Rub{\'\i}}},\ }\bibfield  {title}
  {\bibinfo {title} {Entropic particle transport in periodic channels},\ }\href
  {https://doi.org/10.1016/j.biosystems.2008.03.006} {\bibfield  {journal}
  {\bibinfo  {journal} {BioSystems}\ }\textbf {\bibinfo {volume} {93}},\
  \bibinfo {pages} {16} (\bibinfo {year} {2008}{\natexlab{b}})}\BibitemShut
  {NoStop}%
\bibitem [{\citenamefont {Das}\ \emph {et~al.}(2012{\natexlab{a}})\citenamefont
  {Das}, \citenamefont {Mondal},\ and\ \citenamefont {Ray}}]{Das2012jcp1}%
  \BibitemOpen
  \bibfield  {author} {\bibinfo {author} {\bibfnamefont {M.}~\bibnamefont
  {Das}}, \bibinfo {author} {\bibfnamefont {D.}~\bibnamefont {Mondal}},\ and\
  \bibinfo {author} {\bibfnamefont {D.~S.}\ \bibnamefont {Ray}},\ }\bibfield
  {title} {\bibinfo {title} {Shape fluctuation-induced dynamic hysteresis},\
  }\href {https://doi.org/10.1063/1.3693333} {\bibfield  {journal} {\bibinfo
  {journal} {J. Chem. Phys.}\ }\textbf {\bibinfo {volume} {136}},\ \bibinfo
  {pages} {114104} (\bibinfo {year} {2012}{\natexlab{a}})}\BibitemShut
  {NoStop}%
\bibitem [{\citenamefont {Burada}\ \emph
  {et~al.}(2009{\natexlab{b}})\citenamefont {Burada}, \citenamefont {Schmid},
  \citenamefont {Reguera}, \citenamefont {Rubi},\ and\ \citenamefont
  {H{\"a}nggi}}]{Burada2009epl}%
  \BibitemOpen
  \bibfield  {author} {\bibinfo {author} {\bibfnamefont {P.~S.}\ \bibnamefont
  {Burada}}, \bibinfo {author} {\bibfnamefont {G.}~\bibnamefont {Schmid}},
  \bibinfo {author} {\bibfnamefont {D.}~\bibnamefont {Reguera}}, \bibinfo
  {author} {\bibfnamefont {J.}~\bibnamefont {Rubi}},\ and\ \bibinfo {author}
  {\bibfnamefont {P.}~\bibnamefont {H{\"a}nggi}},\ }\bibfield  {title}
  {\bibinfo {title} {Double entropic stochastic resonance},\ }\href
  {https://doi.org/10.1209/0295-5075/87/50003} {\bibfield  {journal} {\bibinfo
  {journal} {Europhys. Lett.}\ }\textbf {\bibinfo {volume} {87}},\ \bibinfo
  {pages} {50003} (\bibinfo {year} {2009}{\natexlab{b}})}\BibitemShut {NoStop}%
\bibitem [{\citenamefont {Das}\ \emph {et~al.}(2012{\natexlab{b}})\citenamefont
  {Das}, \citenamefont {Mondal},\ and\ \citenamefont {Ray}}]{Das2012pre}%
  \BibitemOpen
  \bibfield  {author} {\bibinfo {author} {\bibfnamefont {M.}~\bibnamefont
  {Das}}, \bibinfo {author} {\bibfnamefont {D.}~\bibnamefont {Mondal}},\ and\
  \bibinfo {author} {\bibfnamefont {D.~S.}\ \bibnamefont {Ray}},\ }\bibfield
  {title} {\bibinfo {title} {Logic gates for entropic transport},\ }\href
  {https://doi.org/10.1103/PhysRevE.86.041112} {\bibfield  {journal} {\bibinfo
  {journal} {Phys. Rev. E}\ }\textbf {\bibinfo {volume} {86}},\ \bibinfo
  {pages} {041112} (\bibinfo {year} {2012}{\natexlab{b}})}\BibitemShut
  {NoStop}%
\bibitem [{\citenamefont {Ai}\ and\ \citenamefont {Liu}(2008)}]{Quan2008jcp}%
  \BibitemOpen
  \bibfield  {author} {\bibinfo {author} {\bibfnamefont {B.~Q.}\ \bibnamefont
  {Ai}}\ and\ \bibinfo {author} {\bibfnamefont {L.~G.}\ \bibnamefont {Liu}},\
  }\bibfield  {title} {\bibinfo {title} {A channel {B}rownian pump powered by
  an unbiased external force},\ }\href {https://doi.org/10.1063/1.2813420}
  {\bibfield  {journal} {\bibinfo  {journal} {J. Chem. Phys.}\ }\textbf
  {\bibinfo {volume} {128}},\ \bibinfo {pages} {024706} (\bibinfo {year}
  {2008})}\BibitemShut {NoStop}%
\bibitem [{\citenamefont {Das}\ \emph {et~al.}(2012{\natexlab{c}})\citenamefont
  {Das}, \citenamefont {Mondal},\ and\ \citenamefont {Ray}}]{Das2012jcp2}%
  \BibitemOpen
  \bibfield  {author} {\bibinfo {author} {\bibfnamefont {M.}~\bibnamefont
  {Das}}, \bibinfo {author} {\bibfnamefont {D.}~\bibnamefont {Mondal}},\ and\
  \bibinfo {author} {\bibfnamefont {D.~S.}\ \bibnamefont {Ray}},\ }\bibfield
  {title} {\bibinfo {title} {Shape change as entropic phase transition: A study
  using {J}arzynski relation},\ }\href@noop {} {\bibfield  {journal} {\bibinfo
  {journal} {J. Chem. Sci.}\ }\textbf {\bibinfo {volume} {124}},\ \bibinfo
  {pages} {21} (\bibinfo {year} {2012}{\natexlab{c}})}\BibitemShut {NoStop}%
\bibitem [{\citenamefont {Nayak}\ \emph {et~al.}(2020)\citenamefont {Nayak},
  \citenamefont {Debnath}, \citenamefont {Das}, \citenamefont {Debnath},\ and\
  \citenamefont {Ghosh}}]{Nayak2020jpcc}%
  \BibitemOpen
  \bibfield  {author} {\bibinfo {author} {\bibfnamefont {S.}~\bibnamefont
  {Nayak}}, \bibinfo {author} {\bibfnamefont {T.}~\bibnamefont {Debnath}},
  \bibinfo {author} {\bibfnamefont {S.}~\bibnamefont {Das}}, \bibinfo {author}
  {\bibfnamefont {D.}~\bibnamefont {Debnath}},\ and\ \bibinfo {author}
  {\bibfnamefont {P.~K.}\ \bibnamefont {Ghosh}},\ }\bibfield  {title} {\bibinfo
  {title} {Escape kinetics of an underdamped colloidal particle from a cavity
  through narrow pores},\ }\href
  {https://doi.org/https://doi.org/10.1021/acs.jpcc.0c04601} {\bibfield
  {journal} {\bibinfo  {journal} {J. Phys. Chem. C}\ }\textbf {\bibinfo
  {volume} {124}},\ \bibinfo {pages} {18747} (\bibinfo {year}
  {2020})}\BibitemShut {NoStop}%
\bibitem [{\citenamefont {Yang}\ \emph {et~al.}(2017)\citenamefont {Yang},
  \citenamefont {Liu}, \citenamefont {Li}, \citenamefont {Marchesoni},
  \citenamefont {Hänggi},\ and\ \citenamefont {Zhang}}]{yang2017pnas}%
  \BibitemOpen
  \bibfield  {author} {\bibinfo {author} {\bibfnamefont {X.}~\bibnamefont
  {Yang}}, \bibinfo {author} {\bibfnamefont {C.}~\bibnamefont {Liu}}, \bibinfo
  {author} {\bibfnamefont {Y.}~\bibnamefont {Li}}, \bibinfo {author}
  {\bibfnamefont {F.}~\bibnamefont {Marchesoni}}, \bibinfo {author}
  {\bibfnamefont {P.}~\bibnamefont {Hänggi}},\ and\ \bibinfo {author}
  {\bibfnamefont {H.~P.}\ \bibnamefont {Zhang}},\ }\bibfield  {title} {\bibinfo
  {title} {Hydrodynamic and entropic effects on colloidal diffusion in
  corrugated channels},\ }\href {https://doi.org/10.1073/pnas.1707815114}
  {\bibfield  {journal} {\bibinfo  {journal} {Proc. Natl. Acad. Sci. U. S. A}\
  }\textbf {\bibinfo {volume} {114}},\ \bibinfo {pages} {9564} (\bibinfo {year}
  {2017})}\BibitemShut {NoStop}%
\bibitem [{\citenamefont {Zhu}\ \emph {et~al.}(2022)\citenamefont {Zhu},
  \citenamefont {Zhou}, \citenamefont {Marchesoni},\ and\ \citenamefont
  {Zhang}}]{Zhu2022prl}%
  \BibitemOpen
  \bibfield  {author} {\bibinfo {author} {\bibfnamefont {Q.}~\bibnamefont
  {Zhu}}, \bibinfo {author} {\bibfnamefont {Y.}~\bibnamefont {Zhou}}, \bibinfo
  {author} {\bibfnamefont {F.}~\bibnamefont {Marchesoni}},\ and\ \bibinfo
  {author} {\bibfnamefont {H.~P.}\ \bibnamefont {Zhang}},\ }\bibfield  {title}
  {\bibinfo {title} {Colloidal stochastic resonance in confined geometries},\
  }\href {https://doi.org/10.1103/PhysRevLett.129.098001} {\bibfield  {journal}
  {\bibinfo  {journal} {Phys. Rev. Lett.}\ }\textbf {\bibinfo {volume} {129}},\
  \bibinfo {pages} {098001} (\bibinfo {year} {2022})}\BibitemShut {NoStop}%
\bibitem [{\citenamefont {Marquet}\ \emph {et~al.}(2002)\citenamefont
  {Marquet}, \citenamefont {Buguin}, \citenamefont {Talini},\ and\
  \citenamefont {Silberzan}}]{Marquet2002prl}%
  \BibitemOpen
  \bibfield  {author} {\bibinfo {author} {\bibfnamefont {C.}~\bibnamefont
  {Marquet}}, \bibinfo {author} {\bibfnamefont {A.}~\bibnamefont {Buguin}},
  \bibinfo {author} {\bibfnamefont {L.}~\bibnamefont {Talini}},\ and\ \bibinfo
  {author} {\bibfnamefont {P.}~\bibnamefont {Silberzan}},\ }\bibfield  {title}
  {\bibinfo {title} {Rectified motion of colloids in asymmetrically structured
  channels},\ }\href {https://doi.org/10.1103/PhysRevLett.88.168301} {\bibfield
   {journal} {\bibinfo  {journal} {Phys. Rev. Lett.}\ }\textbf {\bibinfo
  {volume} {88}},\ \bibinfo {pages} {168301} (\bibinfo {year}
  {2002})}\BibitemShut {NoStop}%
\bibitem [{\citenamefont {Pagliara}\ \emph
  {et~al.}(2014{\natexlab{b}})\citenamefont {Pagliara}, \citenamefont
  {Dettmer}, \citenamefont {Misiunas}, \citenamefont {Lea}, \citenamefont
  {Tan},\ and\ \citenamefont {Keyser}}]{pagliara2014epjs}%
  \BibitemOpen
  \bibfield  {author} {\bibinfo {author} {\bibfnamefont {S.}~\bibnamefont
  {Pagliara}}, \bibinfo {author} {\bibfnamefont {S.~L.}\ \bibnamefont
  {Dettmer}}, \bibinfo {author} {\bibfnamefont {K.}~\bibnamefont {Misiunas}},
  \bibinfo {author} {\bibfnamefont {L.}~\bibnamefont {Lea}}, \bibinfo {author}
  {\bibfnamefont {Y.}~\bibnamefont {Tan}},\ and\ \bibinfo {author}
  {\bibfnamefont {U.~F.}\ \bibnamefont {Keyser}},\ }\bibfield  {title}
  {\bibinfo {title} {Diffusion coefficients and particle transport in synthetic
  membrane channels},\ }\href
  {https://doi.org/https://doi.org/10.1140/epjst/e2014-02324-6} {\bibfield
  {journal} {\bibinfo  {journal} {Eur. Phys. J. Spec. Top.}\ }\textbf {\bibinfo
  {volume} {223}},\ \bibinfo {pages} {3145} (\bibinfo {year}
  {2014}{\natexlab{b}})}\BibitemShut {NoStop}%
\bibitem [{\citenamefont {Borromeo}\ \emph {et~al.}(2006)\citenamefont
  {Borromeo}, \citenamefont {Giusepponi},\ and\ \citenamefont
  {Marchesoni}}]{Borromeo2006pre}%
  \BibitemOpen
  \bibfield  {author} {\bibinfo {author} {\bibfnamefont {M.}~\bibnamefont
  {Borromeo}}, \bibinfo {author} {\bibfnamefont {S.}~\bibnamefont
  {Giusepponi}},\ and\ \bibinfo {author} {\bibfnamefont {F.}~\bibnamefont
  {Marchesoni}},\ }\bibfield  {title} {\bibinfo {title} {Recycled noise
  rectification: An automated maxwell's daemon},\ }\href
  {https://doi.org/10.1103/PhysRevE.74.031121} {\bibfield  {journal} {\bibinfo
  {journal} {Phys. Rev. E}\ }\textbf {\bibinfo {volume} {74}},\ \bibinfo
  {pages} {031121} (\bibinfo {year} {2006})}\BibitemShut {NoStop}%
\bibitem [{\citenamefont {Zhou}(2010)}]{Zhou2010jpcl}%
  \BibitemOpen
  \bibfield  {author} {\bibinfo {author} {\bibfnamefont {H.-X.}\ \bibnamefont
  {Zhou}},\ }\bibfield  {title} {\bibinfo {title} {Diffusion-influenced
  transport of ions across a transmembrane channel with an internal binding
  site},\ }\href {https://doi.org/10.1021/jz100683t} {\bibfield  {journal}
  {\bibinfo  {journal} {J. Phys. Chem. Lett.}\ }\textbf {\bibinfo {volume}
  {1}},\ \bibinfo {pages} {1973} (\bibinfo {year} {2010})}\BibitemShut
  {NoStop}%
\bibitem [{\citenamefont {Licata}\ \emph {et~al.}(2016)\citenamefont {Licata},
  \citenamefont {Mohari}, \citenamefont {Fuqua},\ and\ \citenamefont
  {Setayeshgar}}]{Licata2016bioj}%
  \BibitemOpen
  \bibfield  {author} {\bibinfo {author} {\bibfnamefont {N.~A.}\ \bibnamefont
  {Licata}}, \bibinfo {author} {\bibfnamefont {B.}~\bibnamefont {Mohari}},
  \bibinfo {author} {\bibfnamefont {C.}~\bibnamefont {Fuqua}},\ and\ \bibinfo
  {author} {\bibfnamefont {S.}~\bibnamefont {Setayeshgar}},\ }\bibfield
  {title} {\bibinfo {title} {Diffusion of bacterial cells in porous media},\
  }\href {https://doi.org/https://doi.org/10.1016/j.bpj.2015.09.035} {\bibfield
   {journal} {\bibinfo  {journal} {Biophys J}\ }\textbf {\bibinfo {volume}
  {110}},\ \bibinfo {pages} {247} (\bibinfo {year} {2016})}\BibitemShut
  {NoStop}%
\bibitem [{\citenamefont {Arango-Restrepo}\ \emph {et~al.}(2021)\citenamefont
  {Arango-Restrepo}, \citenamefont {Rubi}, \citenamefont {Kjelstrup},
  \citenamefont {Angelsen},\ and\ \citenamefont {Davies}}]{Arango2021biophys}%
  \BibitemOpen
  \bibfield  {author} {\bibinfo {author} {\bibfnamefont {A.}~\bibnamefont
  {Arango-Restrepo}}, \bibinfo {author} {\bibfnamefont {J.~M.}\ \bibnamefont
  {Rubi}}, \bibinfo {author} {\bibfnamefont {S.}~\bibnamefont {Kjelstrup}},
  \bibinfo {author} {\bibfnamefont {B.~A.~J.}\ \bibnamefont {Angelsen}},\ and\
  \bibinfo {author} {\bibfnamefont {C.~d.~L.}\ \bibnamefont {Davies}},\
  }\bibfield  {title} {\bibinfo {title} {Enhancing carrier flux for efficient
  drug delivery in cancer tissues},\ }\href
  {https://doi.org/10.1016/j.bpj.2021.10.036} {\bibfield  {journal} {\bibinfo
  {journal} {Biophys. J.}\ }\textbf {\bibinfo {volume} {120}},\ \bibinfo
  {pages} {5255} (\bibinfo {year} {2021})}\BibitemShut {NoStop}%
\bibitem [{\citenamefont {Santamaría-Holek}\ \emph {et~al.}(2019)\citenamefont
  {Santamaría-Holek}, \citenamefont {Hernández}, \citenamefont
  {García-Alcántara},\ and\ \citenamefont {Ledesma-Durán}}]{Holek2019cat}%
  \BibitemOpen
  \bibfield  {author} {\bibinfo {author} {\bibfnamefont {I.}~\bibnamefont
  {Santamaría-Holek}}, \bibinfo {author} {\bibfnamefont {S.}~\bibnamefont
  {Hernández}}, \bibinfo {author} {\bibfnamefont {C.}~\bibnamefont
  {García-Alcántara}},\ and\ \bibinfo {author} {\bibfnamefont
  {A.}~\bibnamefont {Ledesma-Durán}},\ }\bibfield  {title} {\bibinfo {title}
  {Review on the macro-transport processes theory for irregular pores able to
  perform catalytic reactions},\ }\href {https://doi.org/10.3390/catal9030281}
  {\bibfield  {journal} {\bibinfo  {journal} {Catalysts}\ }\textbf {\bibinfo
  {volume} {9}},\ \bibinfo {pages} {281} (\bibinfo {year} {2019})}\BibitemShut
  {NoStop}%
\bibitem [{\citenamefont {Carusela}\ and\ \citenamefont
  {Miguel~Rubi}(2021)}]{carusela2021fcdb}%
  \BibitemOpen
  \bibfield  {author} {\bibinfo {author} {\bibfnamefont {M.~F.}\ \bibnamefont
  {Carusela}}\ and\ \bibinfo {author} {\bibfnamefont {J.}~\bibnamefont
  {Miguel~Rubi}},\ }\bibfield  {title} {\bibinfo {title} {Computational model
  for membrane transporters. potential implications for cancer},\ }\bibfield
  {journal} {\bibinfo  {journal} {Front. Cell Dev. Biol.}\ }\textbf {\bibinfo
  {volume} {9}},\ \href
  {https://doi.org/https://doi.org/10.3389/fcell.2021.642665}
  {https://doi.org/10.3389/fcell.2021.642665} (\bibinfo {year}
  {2021})\BibitemShut {NoStop}%
\bibitem [{\citenamefont {Radi}(1998)}]{Radi1998crt}%
  \BibitemOpen
  \bibfield  {author} {\bibinfo {author} {\bibfnamefont {R.}~\bibnamefont
  {Radi}},\ }\bibfield  {title} {\bibinfo {title} {Peroxynitrite reactions and
  diffusion in biology},\ }\href {https://doi.org/10.1021/tx980096z} {\bibfield
   {journal} {\bibinfo  {journal} {Chem. Res. Toxicol}\ }\textbf {\bibinfo
  {volume} {11}},\ \bibinfo {pages} {720} (\bibinfo {year} {1998})}\BibitemShut
  {NoStop}%
\bibitem [{\citenamefont {Ricciardi}(2013)}]{Ricciardi_2013_diffusion}%
  \BibitemOpen
  \bibfield  {author} {\bibinfo {author} {\bibfnamefont {L.~M.}\ \bibnamefont
  {Ricciardi}},\ }\href@noop {} {\emph {\bibinfo {title} {Diffusion Processes
  and Related Topics in Biology}}}\ (\bibinfo  {publisher} {Springer Science \&
  Business Media},\ \bibinfo {year} {2013})\BibitemShut {NoStop}%
\bibitem [{\citenamefont {Kleinfeld}\ \emph {et~al.}(1998)\citenamefont
  {Kleinfeld}, \citenamefont {Storms},\ and\ \citenamefont
  {Watts}}]{Kleinfeld1998biochem}%
  \BibitemOpen
  \bibfield  {author} {\bibinfo {author} {\bibfnamefont {A.~M.}\ \bibnamefont
  {Kleinfeld}}, \bibinfo {author} {\bibfnamefont {S.}~\bibnamefont {Storms}},\
  and\ \bibinfo {author} {\bibfnamefont {M.}~\bibnamefont {Watts}},\ }\bibfield
   {title} {\bibinfo {title} {Transport of long-chain native fatty acids across
  human erythrocyte ghost membranes},\ }\href
  {https://doi.org/https://doi.org/10.1021/bi980301+} {\bibfield  {journal}
  {\bibinfo  {journal} {Biochemistry}\ }\textbf {\bibinfo {volume} {37}},\
  \bibinfo {pages} {8011} (\bibinfo {year} {1998})}\BibitemShut {NoStop}%
\bibitem [{\citenamefont {Berezhkovskii}\ and\ \citenamefont
  {Bezrukov}(2022)}]{Berez2022jcp}%
  \BibitemOpen
  \bibfield  {author} {\bibinfo {author} {\bibfnamefont {A.~M.}\ \bibnamefont
  {Berezhkovskii}}\ and\ \bibinfo {author} {\bibfnamefont {S.~M.}\ \bibnamefont
  {Bezrukov}},\ }\bibfield  {title} {\bibinfo {title} {Intrinsic diffusion
  resistance of a membrane channel, mean first-passage times between its ends,
  and equilibrium unidirectional fluxes},\ }\href
  {https://doi.org/10.1063/5.0082482} {\bibfield  {journal} {\bibinfo
  {journal} {J. Chem. Phys.}\ }\textbf {\bibinfo {volume} {156}},\ \bibinfo
  {pages} {071103} (\bibinfo {year} {2022})}\BibitemShut {NoStop}%
\bibitem [{\citenamefont {Mondal}\ and\ \citenamefont
  {Muthukumar}(2016{\natexlab{a}})}]{Mondal2016jcp1}%
  \BibitemOpen
  \bibfield  {author} {\bibinfo {author} {\bibfnamefont {D.}~\bibnamefont
  {Mondal}}\ and\ \bibinfo {author} {\bibfnamefont {M.}~\bibnamefont
  {Muthukumar}},\ }\bibfield  {title} {\bibinfo {title} {Ratchet rectification
  effect on the translocation of a flexible polyelectrolyte chain},\ }\href
  {https://doi.org/10.1063/1.4961505} {\bibfield  {journal} {\bibinfo
  {journal} {J. Chem. Phys.}\ }\textbf {\bibinfo {volume} {145}},\ \bibinfo
  {pages} {084906} (\bibinfo {year} {2016}{\natexlab{a}})}\BibitemShut
  {NoStop}%
\bibitem [{\citenamefont {Mondal}\ and\ \citenamefont
  {Muthukumar}(2016{\natexlab{b}})}]{Mondal2016jcp2}%
  \BibitemOpen
  \bibfield  {author} {\bibinfo {author} {\bibfnamefont {D.}~\bibnamefont
  {Mondal}}\ and\ \bibinfo {author} {\bibfnamefont {M.}~\bibnamefont
  {Muthukumar}},\ }\bibfield  {title} {\bibinfo {title} {Stochastic resonance
  during a polymer translocation process},\ }\href
  {https://doi.org/10.1063/1.4945559} {\bibfield  {journal} {\bibinfo
  {journal} {J. Chem. Phys.}\ }\textbf {\bibinfo {volume} {144}},\ \bibinfo
  {pages} {144901} (\bibinfo {year} {2016}{\natexlab{b}})}\BibitemShut
  {NoStop}%
\bibitem [{\citenamefont {Muthukumar}(2003)}]{Muthukumar2003jcp}%
  \BibitemOpen
  \bibfield  {author} {\bibinfo {author} {\bibfnamefont {M.}~\bibnamefont
  {Muthukumar}},\ }\bibfield  {title} {\bibinfo {title} {Polymer escape through
  a nanopore},\ }\href {https://doi.org/10.1063/1.1553753} {\bibfield
  {journal} {\bibinfo  {journal} {J. Chem. Phys.}\ }\textbf {\bibinfo {volume}
  {118}},\ \bibinfo {pages} {5174} (\bibinfo {year} {2003})}\BibitemShut
  {NoStop}%
\bibitem [{\citenamefont {Mangeat}\ \emph {et~al.}(2020)\citenamefont
  {Mangeat}, \citenamefont {Gu{\'e}rin},\ and\ \citenamefont
  {Dean}}]{Mangeat2020jcp}%
  \BibitemOpen
  \bibfield  {author} {\bibinfo {author} {\bibfnamefont {M.}~\bibnamefont
  {Mangeat}}, \bibinfo {author} {\bibfnamefont {T.}~\bibnamefont
  {Gu{\'e}rin}},\ and\ \bibinfo {author} {\bibfnamefont {D.}~\bibnamefont
  {Dean}},\ }\bibfield  {title} {\bibinfo {title} {Effective diffusivity of
  {B}rownian particles in a two dimensional square lattice of hard disks},\
  }\bibfield  {journal} {\bibinfo  {journal} {J. Chem. Phys.}\ }\textbf
  {\bibinfo {volume} {152}},\ \href {https://doi.org/10.1063/5.0009095}
  {10.1063/5.0009095} (\bibinfo {year} {2020})\BibitemShut {NoStop}%
\bibitem [{\citenamefont {Wang}\ \emph {et~al.}(2021)\citenamefont {Wang},
  \citenamefont {Chu}, \citenamefont {Tsao},\ and\ \citenamefont
  {Sheng}}]{Wang2021pccp}%
  \BibitemOpen
  \bibfield  {author} {\bibinfo {author} {\bibfnamefont {Z.}~\bibnamefont
  {Wang}}, \bibinfo {author} {\bibfnamefont {K.-C.}\ \bibnamefont {Chu}},
  \bibinfo {author} {\bibfnamefont {H.-K.}\ \bibnamefont {Tsao}},\ and\
  \bibinfo {author} {\bibfnamefont {Y.-J.}\ \bibnamefont {Sheng}},\ }\bibfield
  {title} {\bibinfo {title} {Preferred penetration of active nano-rods into
  narrow channels and their clustering},\ }\href
  {https://doi.org/https://doi.org/10.1039/D1CP01065D} {\bibfield  {journal}
  {\bibinfo  {journal} {Phys. Chem. Chem. Phys.}\ }\textbf {\bibinfo {volume}
  {23}},\ \bibinfo {pages} {16234} (\bibinfo {year} {2021})}\BibitemShut
  {NoStop}%
\bibitem [{\citenamefont {Santamar{\'\i}a-Holek}\ \emph
  {et~al.}(2020)\citenamefont {Santamar{\'\i}a-Holek}, \citenamefont
  {Ledesma-Dur{\'a}n}, \citenamefont {Hern{\'a}ndez}, \citenamefont
  {Garc{\'\i}a-Alc{\'a}ntara}, \citenamefont {Andrio},\ and\ \citenamefont
  {Compa{\~n}}}]{santamaria2020pccp}%
  \BibitemOpen
  \bibfield  {author} {\bibinfo {author} {\bibfnamefont {I.}~\bibnamefont
  {Santamar{\'\i}a-Holek}}, \bibinfo {author} {\bibfnamefont {A.}~\bibnamefont
  {Ledesma-Dur{\'a}n}}, \bibinfo {author} {\bibfnamefont {S.~I.}\ \bibnamefont
  {Hern{\'a}ndez}}, \bibinfo {author} {\bibfnamefont {C.}~\bibnamefont
  {Garc{\'\i}a-Alc{\'a}ntara}}, \bibinfo {author} {\bibfnamefont
  {A.}~\bibnamefont {Andrio}},\ and\ \bibinfo {author} {\bibfnamefont
  {V.}~\bibnamefont {Compa{\~n}}},\ }\bibfield  {title} {\bibinfo {title}
  {Entropic restrictions control the electric conductance of superprotonic
  ionic solids},\ }\href {https://doi.org/https://doi.org/10.1039/C9CP05486C}
  {\bibfield  {journal} {\bibinfo  {journal} {Phys. Chem. Chem. Phys.}\
  }\textbf {\bibinfo {volume} {22}},\ \bibinfo {pages} {437} (\bibinfo {year}
  {2020})}\BibitemShut {NoStop}%
\bibitem [{\citenamefont {Ali}\ \emph {et~al.}(2022)\citenamefont {Ali},
  \citenamefont {Rafeek},\ and\ \citenamefont {Mondal}}]{Ali2022jcp}%
  \BibitemOpen
  \bibfield  {author} {\bibinfo {author} {\bibfnamefont {S.~Y.}\ \bibnamefont
  {Ali}}, \bibinfo {author} {\bibfnamefont {R.}~\bibnamefont {Rafeek}},\ and\
  \bibinfo {author} {\bibfnamefont {D.}~\bibnamefont {Mondal}},\ }\bibfield
  {title} {\bibinfo {title} {Geometric {B}rownian information engine: {U}pper
  bound of the achievable work under feedback control},\ }\href
  {https://doi.org/10.1063/5.0069582} {\bibfield  {journal} {\bibinfo
  {journal} {J. Chem. Phys.}\ }\textbf {\bibinfo {volume} {156}},\ \bibinfo
  {pages} {014902} (\bibinfo {year} {2022})}\BibitemShut {NoStop}%
\bibitem [{\citenamefont {Rafeek}\ \emph {et~al.}(2023)\citenamefont {Rafeek},
  \citenamefont {Ali},\ and\ \citenamefont {Mondal}}]{Rafna2023pre}%
  \BibitemOpen
  \bibfield  {author} {\bibinfo {author} {\bibfnamefont {R.}~\bibnamefont
  {Rafeek}}, \bibinfo {author} {\bibfnamefont {S.~Y.}\ \bibnamefont {Ali}},\
  and\ \bibinfo {author} {\bibfnamefont {D.}~\bibnamefont {Mondal}},\
  }\bibfield  {title} {\bibinfo {title} {Geometric {B}rownian information
  engine: Essentials for the best performance},\ }\href
  {https://doi.org/10.1103/PhysRevE.107.044122} {\bibfield  {journal} {\bibinfo
   {journal} {Phys. Rev. E}\ }\textbf {\bibinfo {volume} {107}},\ \bibinfo
  {pages} {044122} (\bibinfo {year} {2023})}\BibitemShut {NoStop}%
\bibitem [{\citenamefont {Risken}(1996)}]{Risken}%
  \BibitemOpen
  \bibfield  {author} {\bibinfo {author} {\bibfnamefont {H.}~\bibnamefont
  {Risken}},\ }\href@noop {} {\emph {\bibinfo {title} {The Fokker-Planck
  Equation}}}\ (\bibinfo  {publisher} {Springer-Verlag Berlin Heidelberg},\
  \bibinfo {year} {1996})\BibitemShut {NoStop}%
\bibitem [{\citenamefont {Gardiner}(1985)}]{GardiHSM}%
  \BibitemOpen
  \bibfield  {author} {\bibinfo {author} {\bibfnamefont {C.~W.}\ \bibnamefont
  {Gardiner}},\ }\href
  {https://doi.org/https://doi.org/10.1002/bbpc.19850890629} {\emph {\bibinfo
  {title} {Handbook of {S}tochastic {M}ethods for {P}hysics, {C}hemistry and
  the {N}atural Sciences}}},\ Vol.\ \bibinfo {volume} {115}\ (\bibinfo
  {publisher} {Springer-Verlag, Berlin, Heidelberg, New York},\ \bibinfo {year}
  {1985})\BibitemShut {NoStop}%
\bibitem [{\citenamefont {Mondal}(2011)}]{Mondal2011pre}%
  \BibitemOpen
  \bibfield  {author} {\bibinfo {author} {\bibfnamefont {D.}~\bibnamefont
  {Mondal}},\ }\bibfield  {title} {\bibinfo {title} {Enhancement of entropic
  transport by intermediates},\ }\href
  {https://doi.org/10.1103/PhysRevE.84.011149} {\bibfield  {journal} {\bibinfo
  {journal} {Phys. Rev. E}\ }\textbf {\bibinfo {volume} {84}},\ \bibinfo
  {pages} {011149} (\bibinfo {year} {2011})}\BibitemShut {NoStop}%
\bibitem [{\citenamefont {Poland}\ and\ \citenamefont
  {Scheraga}(1966{\natexlab{a}})}]{Poland1966jcp1}%
  \BibitemOpen
  \bibfield  {author} {\bibinfo {author} {\bibfnamefont {D.}~\bibnamefont
  {Poland}}\ and\ \bibinfo {author} {\bibfnamefont {H.~A.}\ \bibnamefont
  {Scheraga}},\ }\bibfield  {title} {\bibinfo {title} {Phase {T}ransitions in
  {O}ne {D}imension and the {H}elix—{C}oil {T}ransition in {P}olyamino
  {A}cids},\ }\href {https://doi.org/10.1063/1.1727785} {\bibfield  {journal}
  {\bibinfo  {journal} {J. Chem. Phys.}\ }\textbf {\bibinfo {volume} {45}},\
  \bibinfo {pages} {1456} (\bibinfo {year} {1966}{\natexlab{a}})}\BibitemShut
  {NoStop}%
\bibitem [{\citenamefont {Poland}\ and\ \citenamefont
  {Scheraga}(1966{\natexlab{b}})}]{Poland1966jcp2}%
  \BibitemOpen
  \bibfield  {author} {\bibinfo {author} {\bibfnamefont {D.}~\bibnamefont
  {Poland}}\ and\ \bibinfo {author} {\bibfnamefont {H.~A.}\ \bibnamefont
  {Scheraga}},\ }\bibfield  {title} {\bibinfo {title} {Occurrence of a {P}hase
  {T}ransition in {N}ucleic {A}cid {M}odels},\ }\href
  {https://doi.org/10.1063/1.1727786} {\bibfield  {journal} {\bibinfo
  {journal} {J. Chem. Phys.}\ }\textbf {\bibinfo {volume} {45}},\ \bibinfo
  {pages} {1464} (\bibinfo {year} {1966}{\natexlab{b}})}\BibitemShut {NoStop}%
\bibitem [{\citenamefont {Bar}\ \emph {et~al.}(2007)\citenamefont {Bar},
  \citenamefont {Kafri},\ and\ \citenamefont {Mukamel}}]{Bar2007prl}%
  \BibitemOpen
  \bibfield  {author} {\bibinfo {author} {\bibfnamefont {A.}~\bibnamefont
  {Bar}}, \bibinfo {author} {\bibfnamefont {Y.}~\bibnamefont {Kafri}},\ and\
  \bibinfo {author} {\bibfnamefont {D.}~\bibnamefont {Mukamel}},\ }\bibfield
  {title} {\bibinfo {title} {Loop {D}ynamics in {DNA} {D}enaturation},\ }\href
  {https://doi.org/10.1103/PhysRevLett.98.038103} {\bibfield  {journal}
  {\bibinfo  {journal} {Phys. Rev. Lett.}\ }\textbf {\bibinfo {volume} {98}},\
  \bibinfo {pages} {038103} (\bibinfo {year} {2007})}\BibitemShut {NoStop}%
\bibitem [{\citenamefont {Kaiser}\ and\ \citenamefont
  {Novotn{\`y}}(2014)}]{Kaiser2014jphysa}%
  \BibitemOpen
  \bibfield  {author} {\bibinfo {author} {\bibfnamefont {V.}~\bibnamefont
  {Kaiser}}\ and\ \bibinfo {author} {\bibfnamefont {T.}~\bibnamefont
  {Novotn{\`y}}},\ }\bibfield  {title} {\bibinfo {title} {Loop exponent in
  {DNA} bubble dynamics},\ }\href
  {https://doi.org/10.1088/1751-8113/47/31/315003} {\bibfield  {journal}
  {\bibinfo  {journal} {J. Phys. A}\ }\textbf {\bibinfo {volume} {47}},\
  \bibinfo {pages} {315003} (\bibinfo {year} {2014})}\BibitemShut {NoStop}%
\bibitem [{\citenamefont {Fogedby}\ and\ \citenamefont
  {Metzler}(2007)}]{Fogedby2007prl}%
  \BibitemOpen
  \bibfield  {author} {\bibinfo {author} {\bibfnamefont {H.~C.}\ \bibnamefont
  {Fogedby}}\ and\ \bibinfo {author} {\bibfnamefont {R.}~\bibnamefont
  {Metzler}},\ }\bibfield  {title} {\bibinfo {title} {{DNA} {B}ubble {D}ynamics
  as a {Q}uantum {C}oulomb {P}roblem},\ }\href
  {https://doi.org/10.1103/PhysRevLett.98.070601} {\bibfield  {journal}
  {\bibinfo  {journal} {Phys. Rev. Lett.}\ }\textbf {\bibinfo {volume} {98}},\
  \bibinfo {pages} {070601} (\bibinfo {year} {2007})}\BibitemShut {NoStop}%
\bibitem [{\citenamefont {Kessler}\ and\ \citenamefont
  {Barkai}(2012)}]{Kessler2012prl}%
  \BibitemOpen
  \bibfield  {author} {\bibinfo {author} {\bibfnamefont {D.~A.}\ \bibnamefont
  {Kessler}}\ and\ \bibinfo {author} {\bibfnamefont {E.}~\bibnamefont
  {Barkai}},\ }\bibfield  {title} {\bibinfo {title} {Theory of {F}ractional
  {L}{\'e}vy {K}inetics for {C}old {A}toms {D}iffusing in {O}ptical
  {L}attices},\ }\href {https://doi.org/10.1103/PhysRevLett.108.230602}
  {\bibfield  {journal} {\bibinfo  {journal} {Phys. Rev. Lett.}\ }\textbf
  {\bibinfo {volume} {108}},\ \bibinfo {pages} {230602} (\bibinfo {year}
  {2012})}\BibitemShut {NoStop}%
\bibitem [{\citenamefont {Kessler}\ and\ \citenamefont
  {Barkai}(2010)}]{Kessler2010prl}%
  \BibitemOpen
  \bibfield  {author} {\bibinfo {author} {\bibfnamefont {D.~A.}\ \bibnamefont
  {Kessler}}\ and\ \bibinfo {author} {\bibfnamefont {E.}~\bibnamefont
  {Barkai}},\ }\bibfield  {title} {\bibinfo {title} {Infinite {C}ovariant
  {d}ensity for {D}iffusion in {L}ogarithmic {P}otentials and {O}ptical
  {L}attices},\ }\href {https://doi.org/10.1103/PhysRevLett.105.120602}
  {\bibfield  {journal} {\bibinfo  {journal} {Phys. Rev. Lett.}\ }\textbf
  {\bibinfo {volume} {105}},\ \bibinfo {pages} {120602} (\bibinfo {year}
  {2010})}\BibitemShut {NoStop}%
\bibitem [{\citenamefont {Lutz}\ and\ \citenamefont
  {Renzoni}(2013)}]{Lutz2013nphys}%
  \BibitemOpen
  \bibfield  {author} {\bibinfo {author} {\bibfnamefont {E.}~\bibnamefont
  {Lutz}}\ and\ \bibinfo {author} {\bibfnamefont {F.}~\bibnamefont {Renzoni}},\
  }\bibfield  {title} {\bibinfo {title} {Beyond {B}oltzmann--{G}ibbs
  statistical mechanics in optical lattices},\ }\href
  {https://doi.org/10.1038/nphys2751} {\bibfield  {journal} {\bibinfo
  {journal} {Nat. Phys.}\ }\textbf {\bibinfo {volume} {9}},\ \bibinfo {pages}
  {615} (\bibinfo {year} {2013})}\BibitemShut {NoStop}%
\bibitem [{\citenamefont {Ali}\ \emph {et~al.}(2024)\citenamefont {Ali},
  \citenamefont {Bauri},\ and\ \citenamefont {Mondal}}]{Ali2024jpcb}%
  \BibitemOpen
  \bibfield  {author} {\bibinfo {author} {\bibfnamefont {S.~Y.}\ \bibnamefont
  {Ali}}, \bibinfo {author} {\bibfnamefont {P.}~\bibnamefont {Bauri}},\ and\
  \bibinfo {author} {\bibfnamefont {D.}~\bibnamefont {Mondal}},\ }\bibfield
  {title} {\bibinfo {title} {Optimizing work extraction in the presence of an
  entropic potential: An entropic stochastic resonance},\ }\href
  {https://doi.org/10.1021/acs.jpcb.3c08066} {\bibfield  {journal} {\bibinfo
  {journal} {J. Phys. Chem. B}\ }\textbf {\bibinfo {volume} {128}},\ \bibinfo
  {pages} {3824} (\bibinfo {year} {2024})}\BibitemShut {NoStop}%
\bibitem [{\citenamefont {Ali}(2025)}]{ali2025ijp}%
  \BibitemOpen
  \bibfield  {author} {\bibinfo {author} {\bibfnamefont {S.~Y.}\ \bibnamefont
  {Ali}},\ }\bibfield  {title} {\bibinfo {title} {Dynamics of {B}rownian
  particles in asymmetric confinement: Insights into entropic stochastic
  resonance},\ }\href
  {https://doi.org/https://doi.org/10.1007/s12648-025-03562-8} {\bibfield
  {journal} {\bibinfo  {journal} {Indian J. Phys.}\ ,\ \bibinfo {pages} {1}}
  (\bibinfo {year} {2025})}\BibitemShut {NoStop}%
\bibitem [{\citenamefont {Xavier}(1994)}]{xavier1994fortran}%
  \BibitemOpen
  \bibfield  {author} {\bibinfo {author} {\bibfnamefont {C.}~\bibnamefont
  {Xavier}},\ }\href@noop {} {\emph {\bibinfo {title} {Fortran 77 and Numerical
  Methods}}}\ (\bibinfo  {publisher} {New Age International},\ \bibinfo {year}
  {1994})\BibitemShut {NoStop}%
\bibitem [{\citenamefont {Box}\ and\ \citenamefont
  {Muller}(1958)}]{Box1958ams}%
  \BibitemOpen
  \bibfield  {author} {\bibinfo {author} {\bibfnamefont {G.~E.~P.}\
  \bibnamefont {Box}}\ and\ \bibinfo {author} {\bibfnamefont {M.~E.}\
  \bibnamefont {Muller}},\ }\bibfield  {title} {\bibinfo {title} {{A} {N}ote on
  the {G}eneration of {R}andom {N}ormal {D}eviates},\ }\href
  {https://doi.org/10.1214/aoms/1177706645} {\bibfield  {journal} {\bibinfo
  {journal} {Ann. Math. Stat}\ }\textbf {\bibinfo {volume} {29}},\ \bibinfo
  {pages} {610 } (\bibinfo {year} {1958})}\BibitemShut {NoStop}%
\end{thebibliography}%
\end{document}